\documentclass[preprint,11pt,a4paper,nopreprintline]{elsarticle}
\usepackage[T1]{fontenc}
\usepackage[utf8]{inputenc}
\usepackage{lmodern,microtype}
\usepackage[a4paper,margin=25mm]{geometry}
\usepackage{amsmath,amssymb,amsthm,mathtools}
\usepackage{booktabs,tabularx,array,enumitem,needspace,etoolbox}
\usepackage{tikz,xurl,xcolor}
\usepackage{flafter,placeins}
\usetikzlibrary{arrows.meta,positioning,calc}
\usepackage[hidelinks,breaklinks=true]{hyperref}
\usepackage[capitalise,noabbrev]{cleveref}
\biboptions{numbers,sort&compress}
\setlist[enumerate]{label=(\arabic*),leftmargin=2em,itemsep=3pt}
\setlist[itemize]{leftmargin=1.5em,itemsep=2pt}
\newtheorem{theorem}{Theorem}[section]
\newtheorem{introresult}{Theorem}

\newtheorem{lemma}[theorem]{Lemma}
\newtheorem{proposition}[theorem]{Proposition}
\newtheorem{corollary}[theorem]{Corollary}
\theoremstyle{definition}
\newtheorem{definition}[theorem]{Definition}
\theoremstyle{remark}
\newtheorem{remark}[theorem]{Remark}
\newcommand{\statementroom}[1]{%
	\BeforeBeginEnvironment{#1}{\Needspace{3\baselineskip}}}
\forcsvlist{\statementroom}{introresult,theorem,lemma,proposition,corollary,definition}
\setlist{topsep=5pt plus 2pt minus 1pt,parsep=0pt,partopsep=0pt}
\AtBeginEnvironment{table}{}
\newcommand{\N}{\mathbb N_0}
\newcommand{\Z}{\mathbb Z}
\newcommand{\E}{\mathbb E}
\newcommand{\Prob}{\mathbb P}
\newcommand{\RDEF}{\mathbf{RDEF}}
\newcommand{\Eq}{\mathsf{Eq}}
\newcommand{\DTIME}{\mathsf{DTIME}}
\newcommand{\NTIME}{\mathsf{NTIME}}
\newcommand{\NPclass}{\mathbf{NP}}
\newcommand{\DSPACE}{\mathsf{DSPACE}}
\newcommand{\NSPACE}{\mathsf{NSPACE}}
\newcommand{\PSPACE}{\mathbf{PSPACE}}
\newcommand{\Pclass}{\mathbf P}
\newcommand{\one}{\mathbf1}
\newcommand{\out}{\operatorname{out}}
\newcommand{\supp}{\operatorname{supp}}
\newcommand{\ceil}[1]{\left\lceil #1\right\rceil}

\newcommand{\otherclockssource}{In a companion paper in preparation we show that the choice of clock changes the answers.}
\hypersetup{pdftitle={Nondeterministic Space and Time from Backward Feedback},pdfauthor={Saman Farhat and Farzad Didehvar}}
\begin{document}
	\begin{frontmatter}
		\title{Nondeterministic Space and Time from Backward Feedback}
		\author[aut]{Saman Farhat}
		\ead{saman.farhat@aut.ac.ir}
		
		\author[aut]{Farzad Didehvar}
		
		\ead{didehvar@aut.ac.ir}
		\affiliation[aut]{organization={Amirkabir University of Technology},city={Tehran},country={Iran}}
		\begin{abstract}
A discovery made late in one run can guide another from its first
step. We let a deterministic program pass a finite register between
runs on the same input. An observer follows one run forward or moves
back to an earlier step of a fresh run, which starts with the register
contents of the step just left. We call this handoff backward feedback.
The step number in view is a random clock whose proposed changes are
independent draws from a fixed distribution with bounded jumps of both
signs and negative mean.

At equilibrium, every stationary distribution of the clock and
register must give the correct bit almost surely. For each fixed
polynomial bound on time and register length, feedback gives
nondeterministic space at that bound. Disabling feedback gives
deterministic time, a strictly smaller class. Their unions are
$\PSPACE$ and $\Pclass$.

Starting from a blank register, an observer must see acceptance by a
deadline counted in clock ticks, with uniformly positive probability on
members and never on nonmembers. Subexponential deadlines then give
exactly the languages determined by a bounded input prefix. Take a
deadline exponential in a fixed polynomial bound. If no initial register
may lead a nonmember to acceptance, the observable class is
nondeterministic time at that bound with feedback and deterministic time
without it. The separation is unconditional for linear bounds, and the
unions are $\NPclass$ and $\Pclass$. If only the blank start must be
safe, the union with feedback is $\PSPACE$, as at equilibrium.

A clock that takes single steps and favors moving back seldom gets
far into a run, so a message ready late takes longer to send. Clearing
a saved count means waiting for all the handoffs needed to fill it again.
\end{abstract}
	\end{frontmatter}
	\section{Introduction}\label{sec:intro}

A testing program may spend an hour finding a sequence of commands that
crashes an application. A copy running on another machine can try that
sequence first. Two agents searching the same map can do something similar.
One sends a route it has found, and the other can start by trying to
shorten it. What can these workers compute if each runs the same program
on the same input and the message must fit in a finite register?

Imagine following the work one run at a time. If we stay with a worker,
we can watch it advance to a later step. Now suppose it sends its
register at step two. Our next observation might be step one of its
successor, which began with that register already in place. The observed
step number has gone back by one. To study handoffs whose timing the
workers cannot choose, we let an external schedule draw a signed change
in step number independently from the same probability distribution
each time. Each draw is a \emph{proposal}. The resulting sequence of
step numbers is our \emph{random clock}. At each \emph{tick} the clock
receives one proposal.

Suppose the proposals are bounded, can go in both directions, and have
negative mean. We read an answer in two ways. At \emph{equilibrium}, we
sample the clock position and the register from a \emph{stationary}
distribution, one that a tick leaves unchanged. From a \emph{blank
start}, at step zero with an all-zero register, we watch until a
deadline counted in ticks. Fix a \emph{polynomial scale}
$s(n)=(n+1)^\alpha$, where $\alpha$ is a positive integer and $n$ is the
input length. For equilibrium readout, $s$ bounds both the ordinary
running time and the register length up to constant factors. For
observation, $2^{O(s)}$ bounds the number of clock ticks. An individual
history may run forever, and the register length may have a different
polynomial bound. With feedback, equilibrium readout gives exactly
nondeterministic space $O(s)$. From a blank start, $2^{O(s)}$ ticks give
nondeterministic time $O(s)$ if no initial message can lead a nonmember
to acceptance. If only the blank start must be safe, they give
polynomial space over all polynomial scales. With transmission
disabled, equilibrium readout and observation in which no initial
message can lead a nonmember to acceptance both give deterministic time
$O(s)$.

\subsection{From a handoff to a clock tick}\label{sec:why}

Fix the input throughout all runs. In the testing example, the fixed
input describes the software under test. The failing command sequence
is the message. A new worker gets that message in its register and
starts with blank work tapes. Its subsequent steps are determined by
the input, the message and the program. We call this sequence of
configurations a \emph{history}. If the program halts, later observations
of the same history see its final configuration.

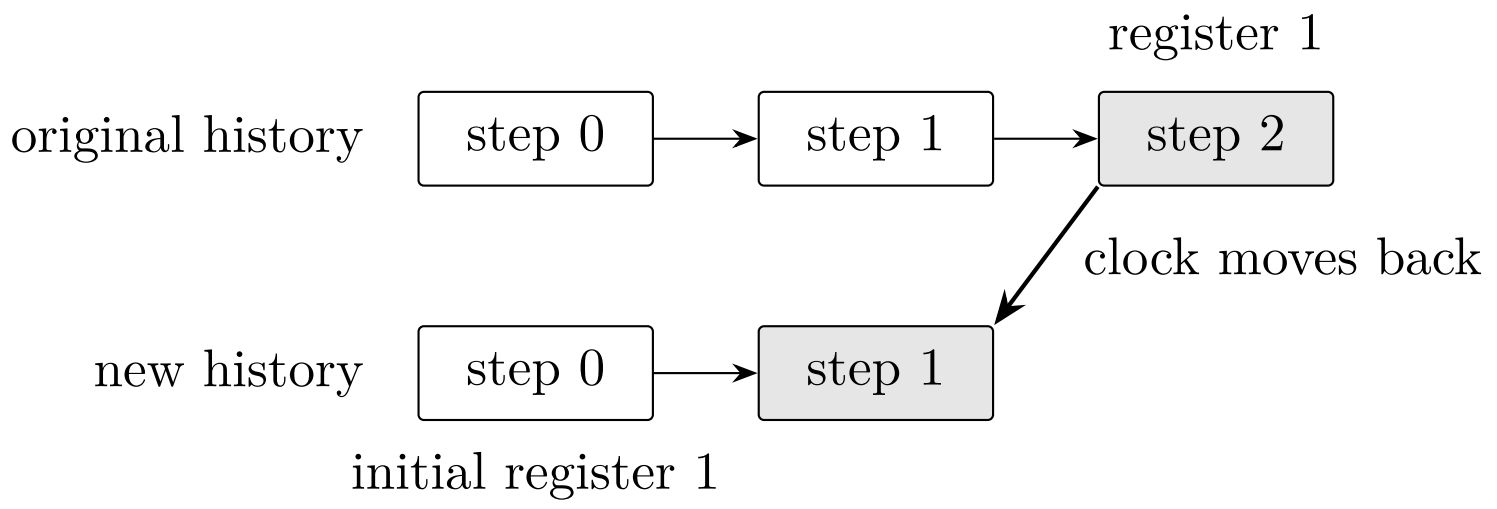
\begin{figure}[htbp]
  \centering
  \begin{tikzpicture}[x=1cm,y=1cm,>=Stealth,font=\small,
    stage/.style={draw,rounded corners=1pt,minimum width=1.55cm,minimum height=.62cm}]
    \node[anchor=east] at (-1,0) {original history};
    \node[anchor=east] at (-1,-1.55) {new history};
    \node[stage] (a0) at (0,0) {step 0};
    \node[stage] (a1) at (2.25,0) {step 1};
    \node[stage,fill=black!10] (a2) at (4.5,0) {step 2};
    \node[stage] (b0) at (0,-1.55) {step 0};
    \node[stage,fill=black!10] (b1) at (2.25,-1.55) {step 1};
    \draw[->] (a0)--(a1);
    \draw[->] (a1)--(a2);
    \draw[->] (b0)--(b1);
    \draw[->,thick] (a2.south west)--(b1.north east)
    node[midway,right=3pt] {clock moves back};
    \node[below=2pt of b0] {initial register $1$};
    \node[above=2pt of a2] {register $1$};
  \end{tikzpicture}
  \caption{The handoff takes the observer from step two of one run to step
    one of another. The second run begins with the register word that
    the first run holds at step two.}
  \label{fig:history-example}
\end{figure}

The clock index, or \emph{height}, counts how many ordinary steps the
worker in view has taken. Each tick selects what we see next. A
positive proposal advances the index within the same history. A negative
proposal that keeps the index at least zero is \emph{executed}: the
current worker sends the word in its register at the step we are
leaving. That word becomes the initial register of a new history, which
we inspect at the proposed earlier index. In \cref{fig:history-example},
a one written by step two of the first run is available before the very
first instruction of the second. We call this operation \emph{backward
feedback}. A proposal of zero, or one that would take the index below
zero, leaves both the history and the index as they were.

A route sent early gives the next worker something to improve.
Waiting another step might uncover a shortcut worth sending. To study
the value of the message when the worker cannot choose the moment to
send it, we draw the proposals independently from the same distribution
at every tick, whatever the input or the register contains. This makes
the clock \emph{autonomous}. The program itself has no random bits.
Randomness enters only through which step we see and when a message is
sent. For example, it might propose one step
forward with probability one third and one step back with probability
two thirds. The register length determines how much survives a handoff.
The negative mean of the proposals is a separate assumption about
timing. It describes a schedule that repeatedly brings early stages into
view, even though a long stretch of forward moves can still reveal a
late discovery. We count ticks separately from the ordinary work a
machine would spend reconstructing the selected step. The
classifications below hold for every fixed distribution with these
properties. Allowing a program or an adversary to choose proposals can
change the computational power (Section~\ref{sec:discussion}).

Even a single bit shows how the two readings differ. Suppose a worker
can report a failure only after step $h$ has been reached. An executed
negative proposal from that height or higher sends a one, and every
later handoff preserves it. With the one-step clock just described,
every starting state admits a sequence of moves that sets the bit.
Every stationary distribution therefore gives the set bit probability
one. Yet an observer who starts with a clear bit must wait until the
clock reaches $h$ and steps back before any worker can inherit the
report.

\subsection{Equilibrium power and its origins}\label{sec:related}

We also use the word \emph{law} for a probability distribution. A
\emph{stationary law} is a distribution of the clock index and the
initial register of the selected history that one tick leaves unchanged.
At equilibrium we draw this pair and read the final bit of the run
started from that register. A stationary law must exist, and every
stationary law must give the right answer with probability one. This
also covers laws supported on states unreachable from a blank start.
Sampling such a law is a resource of the model. For the clocks studied
here, the requirement is equivalent to having the right bit at every
\emph{recurrent} state, one to which the process returns with
probability one when started there (Lemma~\ref{lem:tight}). It is a
safety condition on recurrent states, not a majority vote over runs.

For a bound $s$, write $\DTIME(s)$ and $\NTIME(s)$ for deterministic
and nondeterministic time $O(s)$, and $\DSPACE(s)$ and $\NSPACE(s)$ for
deterministic and nondeterministic work space $O(s)$. Taking unions
over all polynomial bounds gives the classes $\Pclass$, $\NPclass$ and
$\PSPACE$ of polynomial time, nondeterministic polynomial time and
polynomial space.

\begin{introresult}[Equilibrium power]
  Fix a clock law with bounded jumps of both signs and negative mean
  proposal. For each fixed positive integer $\alpha$, put
  $s(n)=(n+1)^\alpha$, where $n$ is the input length. With register
  length $O(s)$ and ordinary running time $O(s)$ from every initial
  register word, equilibrium readout decides exactly $\NSPACE(s)$.
  Disabling transmission gives exactly $\DTIME(s)$, and
  \[
    \DTIME(s)\subsetneq\NSPACE(s).
  \]
\end{introresult}

Deutsch proposed a consistency condition for a state sent around a
closed timelike curve. Aaronson and Watrous showed that polynomial-size
classical and quantum circuits subject to this condition both decide
exactly polynomial space~\citep{Deutsch1991,AaronsonWatrous2009}. They
also obtain polynomial space with a constant-depth circuit inside the
loop~\citep[Section~6.1]{AaronsonWatrous2009}. Later work considers
narrow consistency registers and computations with no bound on register
length or running time~\citep{SayYakaryilmaz2011Narrow,ODonnellSay2014,AaronsonEtAl2024}.
Theorem~A identifies the class at each fixed polynomial scale when
messages must be written sequentially and an external clock decides
when to send them.

The clock itself grew out of Didehvar's work on fuzzy-time computation,
where access to a stage of a computation is described
probabilistically~\citep{Didehvar2023,Didehvar2026}. In our earlier
model, a backward move revisited an earlier step of the same run. For
bounded jumps with negative mean and polynomially many ticks, the
observable languages were exactly those determined by a bounded input
prefix~\citep[Corollary~4.3]{DidehvarFarhat2026RandomClocks}. For a
single history the same conclusion with subexponentially many ticks follows
from Lemmas~4.2 and~2.14 there. Here a backward move can change which
run the observer follows.

Transmission already matters at linear scale. With feedback we
obtain the context-sensitive languages, up to the convention for the
empty word~\citep{Kuroda1964}. With transmission disabled each initial
register word persists. Equilibrium correctness must then hold for
every word, so running the ordinary program once with a blank register
suffices, and the class is deterministic linear time. Over all
polynomial scales the two equilibrium classes are $\PSPACE$ with
feedback and $\Pclass$ without it. Whether they differ is open.

Why can a deterministic program realize a nondeterministic space
computation? A worker prepares one possible successor configuration
at an early step and another at a later step. The clock can transmit
either one. To keep a handoff halfway through a write from exposing
an incomplete message, the program fills a spare register slot before
switching a pointer. A one-bit flag protects the chosen successor during
the return to smaller indices. Building on the reset construction of
Aaronson and Watrous, whose lower-bound lemma they credit to Lance
Fortnow~\citep[Section~3.1, Lemma~2]{AaronsonWatrous2009}, we arrange
that every message can reach a region where the correct answer persists.
A state outside that region with the wrong answer has a positive chance
of entering the region and never returning. It therefore cannot carry
stationary mass.

For the upper bound, the difficulty is that the clock can visit
arbitrarily large indices. The negative mean of the proposals ensures
that stationary laws exist. To test which states can carry their mass, it suffices to examine
return paths between states in a finite strip near zero. Beyond the
halting index the register has stopped changing within each history, so
one counter can describe how far a path has climbed. A short-path
theorem for one-counter systems makes these return tests possible
in nondeterministic space $O(s)$~\citep{ChistikovEtAl2019}.

\subsection{What a blank start reveals}

An ordinary program decides parity by scanning its binary input once.
Why can it take exponentially many clock ticks to see the result? Each
new worker starts at the beginning of the same input. To inspect a
distant symbol, some worker must reach it. Passing a message can save
what one worker has already found, which raises the question of how
much an observer can learn across many short runs.

Starting with a blank register at index zero, the observer checks the
control state before the first tick and after each tick. The deadline
is a \emph{tick budget} $Q(n)$ that depends only on the input length.
On a member the observer must see acceptance by the deadline with
probability at least a fixed $\beta>0$. Under \emph{initialized
soundness}, a nonmember must never be observed accepting from this
blank start. Under \emph{whole-register soundness}, a run on a
nonmember must never enter an accepting state, whatever initial message
it is given.

\begin{introresult}[Observation from a blank start]
  For the clocks of Theorem~A, let $Q(n)\geq n+1$ be a nondecreasing
  tick budget. If $\log Q(n)=o(n)$, both soundness rules give exactly
  the languages determined by a bounded input prefix. Fix $s(n)=(n+1)^\alpha$ for a positive integer $\alpha$, and allow $2^{Cs(n)}$ ticks for
  some constant $C$. Under whole-register soundness this gives exactly
  $\NTIME(s)$ with feedback and $\DTIME(s)$ with transmission disabled.
  Under initialized soundness, the class with feedback lies between
  $\DSPACE(s)$ and $\NSPACE(s)$.
\end{introresult}

When $\log Q(n)=o(n)$, the clock height bound gives each member
an accepting path that reaches only $o(n)$ steps into any run. Shorten
the input to the prefix such a path can read, then replay the same
proposals. The new workers inherit the same register prefix at every
handoff, so the shorter input is also observed accepting, and soundness
makes it a member. A path of minimum accepting height retains that
minimum after shortening, since any lower path could be replayed on the
original input. Applying the height bound to the shorter word now
forces a fixed bound on this minimum for every member. Parity has no
such prefix bound, since appending a one always changes the answer.

With a tick budget exponential in $s$, each member has an accepting
observation at height $O(s)$. Only $O(s)$ initial register symbols
can have been read by then. Under whole-register soundness, guessing
these symbols and simulating the selected run gives a nondeterministic
$O(s)$-time decision procedure. Conversely, a worker can check one
candidate certificate and pass the next candidate to its successor.
The visible verdict belongs to the candidate just checked. Enough
handoffs take the observer through every candidate. Thus feedback
realizes nondeterministic space at equilibrium and nondeterministic
time under whole-register soundness. At linear scale the observation
classes with and without feedback therefore differ unconditionally,
since deterministic and nondeterministic linear time differ by the
multitape theorem of Paul, Pippenger, Szemer\'edi and
Trotter~\citep{PaulPippengerSzemerediTrotter1983}.

Initialized soundness gives a different use for the same message. A
worker can pass the current configuration of a deterministic
space-bounded computation to the next worker, which advances it by
one step. Starting from a blank register ensures that every inherited
configuration is genuine. This gives the lower bound $\DSPACE(s)$ in
Theorem~B. For the upper bound, a nondeterministic machine replays an
accepting observation of height $O(s)$ and stores only the part of the
register that such a run can reach (Proposition~\ref{prop:init-pspace}).

With exponential tick budgets, taking the union over polynomial scales
gives $\Pclass$ without feedback and $\NPclass$ with feedback under
whole-register soundness. With feedback and soundness required only
from the blank start, the same union is $\PSPACE$, as it is for
equilibrium readout.

\subsection{Waiting for a handoff and keeping progress}\label{sec:intro-costs}

The worker who reports a failure needs one handoff after the report is
ready. A worker keeping a count may need many. Suppose the clock moves
one step at a time and is more likely to step back than forward. It
spends much of its time near the beginning of a run, so having a
message ready even a few steps later can multiply the wait before it
is sent. If we want the same chance of seeing all the handoffs we
need, the time we must allow grows in proportion to their number.

Now imagine that each successful handoff adds one to a stored count
until the count is full. Occasionally the count is wiped out and the
next worker has to start it again from zero while the clock keeps
moving. When handoffs take longer or more of them are needed, the
count has to be wiped out less often for us to have the same chance
of finding it full. Section~\ref{sec:costs} gives precise bounds for
both the wait and how often the count can be lost.

\section{The backward-feedback model}\label{sec:model}
	
	The model has two kinds of steps. An ordinary program step updates the
	configuration within the current history. At a clock tick, the
	clock selects an index. Depending on the selected index, a register may
	be transmitted and the current history replaced.
	
	\paragraph{Notation}
	We write $\N=\{0,1,2,\ldots\}$ for the natural numbers and $\Z$ for the
	integers. We write $\Prob$ for probability and $\E$ for expectation.
	The indicator $\one\{E\}$ equals one if the event $E$ occurs and zero
	otherwise. For a language $L$, we write $\one_L(x)=\one\{x\in L\}$.
	An event occurs \emph{almost surely} if it has probability one.
	We use \emph{law} as a synonym for probability distribution.
	
	The word $0^m$ consists of $m$ zero symbols. We write $g=O(f)$ if
	$|g|$ is eventually bounded by a constant multiple of $|f|$, and
	$g=o(f)$ if $g/f$ tends to zero. All logarithms have base two unless
	stated otherwise. The symbols $\lfloor a\rfloor$ and $\lceil a\rceil$
	denote the floor and ceiling of $a$, respectively. We take
	$\inf\varnothing=+\infty$.
	
	\subsection{Logical histories and the message register}
	
	Fix a finite input alphabet $\Sigma$ and a finite register alphabet
	$\Gamma$ containing $0$ and $1$. For an input word $x\in\Sigma^*$,
	write $n=|x|$. We use the usual binary input alphabet for strict class
	separations and completeness results. Symbols from any larger fixed
	alphabet can be encoded by binary blocks of constant length.
	The register holds $\ell(n)$ symbols, so its possible
	initial contents are the words in $R_n=\Gamma^{\ell(n)}$.
	
	An ordinary program is a deterministic multitape Turing machine with a
	read-only input tape, work tapes, and a read-write register tape. The
	same finite program serves all input lengths. Given $x$ and an initial
	register word $r\in R_n$, the program determines a sequence of
	configurations, which we call its \emph{logical history}. At
	configuration index zero, the input tape holds $x$, the register holds
	$r$, the work tapes are blank, and each head is at its leftmost symbol
	position.
	
	Each transition reads the symbols under the heads and writes to the
	writable tapes before moving each head by at most one cell. We number
	symbol positions starting at one. A head can therefore reach cell $j+1$
	at configuration index $j$, but its first opportunity to read that cell
	is transition $j+1$. Endmarkers delimit the input and the register.
	The machine can inspect an endmarker only when its head reaches it.
	The machine receives no separate copy of the input length.
	
	Our machine is therefore an ordinary multitape Turing machine, with
	the register as one more tape. Every tape is read sequentially, so in
	$j$ steps a run sees at most the first $j$ cells of the input and of
	the register. The results of Section~\ref{sec:observation} use this.
	
	We use the standard multitape complexity classes with constant-factor
	resource bounds. The class $\DTIME(s)$ consists of languages decidable
	in deterministic time $O(s)$. We write $\NTIME(s)$ for the corresponding
	nondeterministic class. Every branch takes $O(s)$ steps, and a word is
	accepted exactly when some branch accepts. The classes $\DSPACE(s)$ and $\NSPACE(s)$
	use $O(s)$ work space on deterministic and nondeterministic machines,
	respectively. At each nonhalting step, a deterministic machine has a
	single next configuration, while a nondeterministic machine may choose
	from finitely many. Space counts the work-tape cells used, including
	those that store auxiliary descriptions. It excludes the read-only
	input tape. When encoding a configuration, we include the input-head
	position. In resource bounds with $\log s$ in the denominator, we use
	$\log_2(\max\{2,s\})$, which equals one for $0\leq s\leq 2$.
	
	The classes at a fixed scale depend on the machine model. Two
	results we use are proved for multitape Turing machines: the time--space
	simulation behind the strict inclusion in Theorem~\ref{thm:main}, and
	the linear-time separation~\citep{PaulPippengerSzemerediTrotter1983}
	used for Theorem~\ref{thm:observation} at linear scale. For
	random-access machines, the separation of deterministic from
	nondeterministic linear time is open. The unions $\Pclass$,
	$\NPclass$ and $\PSPACE$ do not depend on this choice, since
	multitape Turing machines and random-access machines with logarithmic
	cost simulate each other with polynomial slowdown.
	
	For equilibrium computation, we use a resource scale $s(n)$ to specify
	both the register length $\ell(n)$ and the ordinary halting-time bound
	$T(n)$. Fix an integer $\alpha\geq 1$ and positive integers $c,d$, and set
	\begin{equation}\label{eq:resource}
		\begin{aligned}
			s(n) &= (n+1)^\alpha,\\
			\ell(n) &= c\,s(n), \qquad T(n) = d\,s(n).
		\end{aligned}
	\end{equation}
	We may increase $c$ and $d$ when choosing bounds for a construction.
	Once chosen, they are fixed across all input lengths. The same finite
	program runs at every length and receives no advice that depends on $n$.
	
	Fix an input $x$ of length $n$, and write $s=s(n)$, $\ell=\ell(n)$,
	and $T=T(n)$. For every initial register word $r\in R_n$, the program
	must halt within $T$ ordinary steps and return a bit $\out_x(r)$.
	Thus $\out_x$ is defined on all of $R_n$. The equilibrium correctness
	condition below specifies how to evaluate this bit when the initial
	register word follows a stationary law.
	
	For each $i\in\N$, let $C_i^{x,r}$ denote the configuration at logical
	index $i$ in the history starting from $(x,r)$. After the program halts,
	we repeat its final configuration at every subsequent index. Define
	\[
	F_{x,i}(r)=\text{the register word in }C_i^{x,r}.
	\]
	Since every history has halted by index $T$, we have $F_{x,i}=F_{x,T}$
	for every $i\geq T$. Direct simulation computes $F_{x,i}(r)$ by
	simulating at most $\min\{i,T\}$ ordinary steps and computes $\out_x(r)$
	by running to the halt. Both computations use $O(s)$ space, since the
	register holds $\ell$ symbols and each work-tape head visits at most
	$T+1$ cells.
	
	\subsection{An autonomous clock with feedback}
	
	Fix an input $x$ of length $n$. At tick $t\in\N$, the joint state is
	$(X_t,R_t)\in\N\times R_n$, where $X_t$ is the clock index and $R_t$
	is the persistent register word. The tick count $t$ records how many
	clock updates have occurred. The index $X_t$ specifies a position in
	the logical history with initial register word $R_t$.
	
	At each tick $t$, the clock draws an integer proposal $K_t$ from a
	fixed probability law $\mu$ on $\Z$, independently of the current state
	and all earlier proposals. Given $(X_t,R_t)=(i,r)$, the next state is
	\begin{equation}\label{eq:joint}
		(X_{t+1},R_{t+1})=
		\begin{cases}
			(i,r),                 & i+K_t<0,\\
			(i+K_t,r),             & K_t\geq 0,\\
			(i+K_t,F_{x,i}(r)),    & K_t<0 \text{ and } i+K_t\geq 0.
		\end{cases}
	\end{equation}
	A proposal that would take the clock below zero causes a \emph{hold}:
	the clock stays at $i$ and the register keeps its value. Only an
	executed backward move transmits a message. The message is the entire
	register word $F_{x,i}(r)$ at the \emph{departure index} $i$ of the
	current history. This word serves as the initial register word of the
	history selected for the next tick. The clock then indexes that history
	at the arrival position $i+K_t$.
	
	The independent proposals make the joint process a Markov chain. Its
	\emph{transition kernel} gives the distribution of the next state for
	each current state. The clock is autonomous because the distribution of
	$X_{t+1}$ given $(X_t,R_t)=(i,r)$ depends only on $i$. The clock and
	register can still be correlated. The \emph{clock marginal} of a joint
	law is the distribution of $X_t$ alone.
	
	An executed backward move uses two operations supplied by the model.
	It accesses the register at a specified index of the history determined
	by $(x,r)$, and it transmits that word in full. When counting clock
	ticks, we charge one tick for this update. The ordinary steps a
	conventional machine would use to simulate $C_i^{x,r}$ are a separate
	cost. When a history halts within $T(n)$ steps, the register at
	index $i$ can be reconstructed by simulating at most $\min\{i,T(n)\}$
	ordinary steps. The observation deadlines in Section~\ref{sec:observation}
	count clock ticks. Remark~\ref{rem:work} shows that charging these
	ordinary steps as well does not change the observation classes.
	
	The support of the proposal law is the set of possible jumps,
	$\supp\mu=\{k\in\Z:\mu(k)>0\}$. Throughout the main equilibrium
	analysis, we fix a law $\mu$ with bounded jumps in both directions
	and negative mean. Thus, for a proposal $K$ with law $\mu$, there are
	positive integers $B,a,b$ such that
	\begin{equation}\label{eq:clock-assumption}
		\supp\mu\subseteq[-B,B],\qquad \E K<0,\qquad
		\mu(a)>0,\qquad \mu(-b)>0.
	\end{equation}
	We call the condition $\E K<0$ \emph{negative drift}. It concerns the
	proposed jump. Near zero, a hold can make the actual displacement
	differ from the proposal.
	
	The same law $\mu$ applies to every input, so its finite support can
	be built into the program. The proof of the definite-answer
	classification uses only the paths allowed by this support and the
	negative drift assumption. It requires no algorithm for computing the
	individual probabilities $\mu(k)$.
	
	\paragraph{Paths and classes}
	A \emph{supported path} is a finite sequence of transitions, each with
	positive probability. A state can \emph{reach} another if a supported
	path leads from the first to the second. Two states \emph{communicate}
	if each can reach the other. A \emph{communicating class} is a maximal
	set of mutually communicating states. A class is \emph{closed} if every
	supported transition from a state in the class stays in the class.
	A chain is \emph{irreducible} if all its states communicate.
	
	For a chain started at a given state, the \emph{first return time} is
	the first positive time at which the chain visits that state again.
	We take this time to be $+\infty$ if the chain never returns. The state
	is \emph{recurrent} if its first return time is finite almost surely,
	and \emph{transient} if it has positive probability of never returning.
	It is \emph{positive recurrent} if its expected first return time is
	finite. A \emph{recurrent class} is a communicating class whose states
	are recurrent.
	
	\subsection{Definite equilibrium readout}
	
	A probability law $\zeta$ on $\N\times R_n$ is \emph{stationary} if
	applying one tick of~\eqref{eq:joint} to a state drawn from $\zeta$
	again gives a state with law $\zeta$. The clock index and the register
	word may be correlated under $\zeta$.
	
	\begin{definition}[Definite equilibrium computation]\label{def:eq}
		A program decides a language $L$ by \emph{definite equilibrium readout}
		if, for every input $x$, the chain~\eqref{eq:joint} has at least one
		stationary probability law and every stationary probability law $\zeta$
		satisfies
		\[
		\zeta\bigl\{(i,r):\out_x(r)=\one_L(x)\bigr\}=1.
		\]
		
		For a fixed proposal law $\mu$ and resource scale $s$, the class
		$\Eq_\mu[O(s)]$ consists of the languages decided in this sense by a
		uniform program satisfying~\eqref{eq:resource} for some positive
		integers $c,d$. These constants may depend on the program and on $\mu$
		and are fixed across all inputs. We write $\Eq_\mu^{\rm off}[O(s)]$
		for the corresponding class when the register retains its value on
		every clock tick, including executed backward moves.
	\end{definition}
	
	By Lemma~\ref{lem:stationarysupport}, the condition holds exactly when
	every recurrent state $(i,r)$ of the chain satisfies
	$\out_x(r)=\one_L(x)$. Section~\ref{sec:classification} works with this
	form. Requiring at least one stationary law makes the correctness condition
	nonvacuous. Correctness must hold even for stationary laws supported
	on classes unreachable from the blank-start state $(0,0^{\ell(n)})$.
	Given a sample $(i,r)$, equilibrium readout returns $\out_x(r)$, the
	final answer of the logical history with initial register word $r$.
	This history may still be running at index $i$. The definition treats
	sampling from a stationary law as an ideal resource and imposes no
	convergence-rate requirement from any initial state.
	
	Section~\ref{sec:observation} defines \emph{initialized observation}.
	It uses the same sequential access conventions and reads the control
	state visible at the current logical index.
	
	\subsection{A one-bit message}\label{sec:latch}
	
	Consider a register that stores a single bit $u\in\{0,1\}$. The bit may
	be encoded in several register symbols. We call the decoded bit the
	\emph{interpreted message}. The update map sets the bit to one,
	$f(0)=f(1)=1$, and the final readout returns its value, $\gamma(u)=u$.
	
	The program constructed in Lemma~\ref{lem:compiler} has a threshold $h\geq 1$.
	On an executed backward move from index $i<h$, the transmitted word
	encodes the current bit. On an executed backward move from $i\geq h$,
	it encodes $f(u)=1$. Thus, once the interpreted message is one, it
	remains one.
	
	Take a nearest-neighbor clock with both proposals $+1$ and $-1$ in its
	support. From any index $i$, a sequence of $\max\{0,h-i\}$ upward moves
	followed by one downward move sets the bit to one. This sequence has
	positive probability. Under the negative drift assumption,
	Lemma~\ref{lem:tight} gives the existence of a stationary law. Since every
	state can reach a state with message one and every subsequent tick
	preserves that bit, every stationary law assigns probability one to
	states with $\out_x(r)=1$. The definite equilibrium answer is therefore
	one.
	
	From the blank start, the stored bit remains zero until the first
	downward move whose departure index is at least $h$. Observing
	acceptance also requires the clock to visit an index where the
	selected logical history has an accepting control state.
	The latch illustrates the two readouts, but its constant answer gives
	no complexity lower bound.

	Section~\ref{sec:costs} computes how long the clock takes to send the
	first update and extends the calculation to a counter that needs
	several updates.
	
	\section{The computational power of backward feedback}\label{sec:classification}
	
	At the same resource scale $s$, enabling backward feedback changes the
	equilibrium class from deterministic time $O(s)$ to nondeterministic
	space $O(s)$.
	
	\begin{theorem}[Equilibrium classification]\label{thm:main}
		For every proposal law $\mu$ satisfying~\eqref{eq:clock-assumption}
		and every fixed positive integer $\alpha$, let $s(n)=(n+1)^\alpha$.
		Then
		\[
		\Eq_\mu[O(s)]=\NSPACE(s),\qquad
		\Eq_\mu^{\rm off}[O(s)]=\DTIME(s)\subsetneq\NSPACE(s).
		\]
	\end{theorem}
	
	For the upper bound with feedback, we reduce the analysis of the
	unbounded clock to a finite test at indices near zero. For the lower
	bound, we encode a space-bounded computation in maps defined on every
	register word and implement those maps by sequential writes. The
	technical proofs appear in \ref{app:compiler}
	and~\ref{app:boundary}.
	
	\subsection{A finite boundary controls the unbounded clock}
	
	Write $v\leadsto w$ if there is a finite supported path from joint state
	$v$ to joint state $w$. We allow paths of length zero, so $v\leadsto v$.
	Let $b$ be the least positive integer with $\mu(-b)>0$. Define the
	\emph{boundary strip} by
	\[
	\mathcal B_n=\{0,\ldots,b-1\}\times R_n,
	\]
	and call its elements \emph{boundary states}. Every joint state can
	reach the strip. Starting at clock index $i$, a sequence of
	$\lfloor i/b\rfloor$ proposals of size $-b$ takes the clock to index
	$i\bmod b$. Each move in this sequence is executed, and the register
	may change along the way. For a nearest-neighbor clock, the strip is
	$\{0\}\times R_n$.
	
	The recurrence test below compares boundary states. The paths used in
	the test may pass through states at any height. Write $\Prob_i$ and
	$\E_i$ for probability and expectation for the clock started at index
	$i$. Its law depends only on $i$.
	
	\begin{lemma}[Stationary laws and the boundary test]\label{lem:tight}\label{lem:boundary}\label{lem:stationarysupport}
		Under~\eqref{eq:clock-assumption}, there are constants $\theta>0$ and
		$A<\infty$, depending only on $\mu$, such that for every initial clock
		index $i\in\N$,
		\begin{equation}\label{eq:expbound}
			\E_i e^{\theta X_t}\leq e^{\theta i}+A\qquad(t\geq 0).
		\end{equation}
		For every input $x$, the joint chain has a stationary probability law.
		Every closed communicating class is positive recurrent and supports a
		stationary probability law.
		
		A boundary state $v$ is recurrent if and only if
		\begin{equation}\label{eq:boundarycriterion}
			\text{for every }w\in\mathcal B_n,\quad
			v\leadsto w\ \Longrightarrow\ w\leadsto v.
		\end{equation}
		At least one boundary state is recurrent, and every recurrent class
		contains a boundary state.
		
		For any bit $q\in\{0,1\}$, every stationary law assigns probability one
		to the event $\out_x(r)=q$ if and only if $\out_x(r)=q$ at every
		recurrent joint state $(i,r)$.
	\end{lemma}
	
	The exponential moment bound~\eqref{eq:expbound} controls the probability
	of large clock indices uniformly in time. Since $R_n$ is finite, the
	time-averaged joint laws have a convergent subsequence whose limit is
	stationary. Starting the chain in a closed communicating class gives
	a stationary law supported on that class and hence proves that the class
	is positive recurrent.
	
	To see why the boundary test suffices, suppose that every boundary state
	reachable from $v$ can return to $v$. Take any state $u$ reachable
	from $v$. By descending in steps of size $-b$, it can reach some
	$w\in\mathcal B_n$. Since $v\leadsto u\leadsto w$, the hypothesis gives
	$w\leadsto v$. Thus every state reachable from $v$ can return to $v$,
	so the reachable states form a closed communicating class. Conversely,
	if $v$ is recurrent, its communicating class is closed, so every state
	reachable from $v$ can return to it. \ref{app:boundary} proves
	the moment bound and the remaining assertions about the joint chain.
	
	\begin{lemma}[Boundary reachability in space $O(s)$]\label{lem:reach}
		For a program satisfying~\eqref{eq:resource}, the predicate
		$v\leadsto w$ for $v,w\in\mathcal B_n$ belongs to $\NSPACE(s)$.
	\end{lemma}
	
	At every departure index $i\geq T$, an executed backward move uses the
	same register map, $F_{x,T}$. This lets us represent the unbounded part
	of the clock by a finite control and one nonnegative counter. A
	short-path theorem for one-counter systems bounds the length of a
	witnessing path by a polynomial in the number of controls. There are
	at most $2^{O(s)}$ controls, so a witness of length at most $2^{O(s)}$
	suffices whenever a path exists. A nondeterministic machine can guess
	such a path one transition at a time, using $O(s)$ bits for the current
	control, counter value, and step count. \ref{app:boundary}
	gives the encoding and verifies the transitions at counter value zero.
	
	\subsection{The space upper bound}
	
	To decide membership in $L$, it suffices to find one recurrent boundary
	state and read its final bit. The main step is to test recurrence using
	the boundary reachability procedure.
	
	\begin{proposition}\label{prop:upper}
		Under~\eqref{eq:clock-assumption}, $\Eq_\mu[O(s)]\subseteq\NSPACE(s)$.
	\end{proposition}
	
	\begin{proof}
		Fix an equilibrium decider for $L$ and an input $x$ of length $n$.
		By Lemma~\ref{lem:reach}, we can test $v\leadsto w$ for boundary states
		$v,w$ in nondeterministic space $O(s)$. Since $s(n)\geq n+1$, the
		Immerman--Szelepcs\'enyi theorem~\citep{Immerman1988,Szelepcsenyi1988}
		also gives a nondeterministic $O(s)$-space test for $w\not\leadsto v$.
		
		By~\eqref{eq:boundarycriterion}, a boundary state $v$ is transient
		exactly when some boundary state $w$ satisfies both $v\leadsto w$ and
		$w\not\leadsto v$. To test for transience, guess $w$ and run the two
		tests in sequence, accepting only if both accept. This transience
		test runs in nondeterministic space $O(s)$, so the
		Immerman--Szelepcs\'enyi theorem, applied once more, gives a
		nondeterministic $O(s)$-space test for recurrence.
		Each boundary state has an $O(s)$-bit description. The procedures use
		the original input $x$ and the stored descriptions of $v$ and $w$,
		with the joint graph represented implicitly throughout.
		
		Lemma~\ref{lem:tight} guarantees that a recurrent boundary state exists.
		Since the program decides $L$ by definite equilibrium readout, the same
		lemma gives $\out_x(r)=\one_L(x)$ at every recurrent state $(i,r)$.
		Consequently,
		\[
		x\in L
		\quad\Longleftrightarrow\quad
		\text{there exists a recurrent }(i,r)\in\mathcal B_n
		\text{ with }\out_x(r)=1.
		\]
		Guess an index $0\leq i<b$ and a word $r\in R_n$, test $(i,r)$ for
		recurrence, and simulate the ordinary program on $(x,r)$ to obtain its
		final bit. Accept if the recurrence test accepts and the bit is one.
		The simulation uses $O(s)$ space, so the entire procedure runs in
		nondeterministic space $O(s)$.
	\end{proof}
	
	The drift assumption ensures positive recurrence of the closed
	communicating classes. Each recurrent state therefore has positive mass
	under some stationary law. A wrong bit at even one such state violates
	definite equilibrium correctness. For bounded-error readout, the
	probability of a wrong answer depends on the total stationary mass of
	the states carrying that bit. Support alone does not give this bound.
	
	\subsection{Total message maps for a space computation}
	
	A stationary law may be supported on registers that are unreachable
	from the blank start. The lower-bound construction must therefore work
	from every message, including malformed strings and encodings of
	configurations that the simulated machine never visits. We construct
	two maps, each defined on every message. From any message, some finite
	sequence of map applications reaches a distinguished message $z_x$.
	From $z_x$, every sequence preserves the correct answer bit.
	
	We adapt the use of configurations and resets from the classical
	consistency construction of Aaronson and Watrous, who credit their
	lower-bound lemma to Lance Fortnow~\citep[Section~3.1, Lemma~2]{AaronsonWatrous2009}.
	Here closure under complementation gives nondeterministic $O(s)$-space
	machines for both $L$ and its complement. Lemma~\ref{lem:gating} and
	the compiler of Lemma~\ref{lem:compiler} then implement the two maps
	in a single sequential program.
	
	\begin{lemma}[Reachable correct region]\label{lem:maps}
		For every $L\in\NSPACE(s)$, there is a message length $m(n)=O(s(n))$
		with the following property. For each input $x$ of length $n$, let
		$M_n=\{0,1\}^{m(n)}$. There are two total maps $f_0,f_1\colon M_n\to M_n$,
		a readout $\gamma\colon M_n\to\{0,1\}$, and a distinguished message
		$z_x\in M_n$ such that
		\begin{enumerate}
			\item From each $u\in M_n$, some finite sequence of applications of
			$f_0$ and $f_1$ reaches $z_x$.
			\item Every $u\in M_n$ reachable from $z_x$ by such a sequence,
			including the empty sequence, satisfies $\gamma(u)=\one_L(x)$.
			\item Given $x$ and $u\in M_n$, the values $f_0(u)$, $f_1(u)$, and
			$\gamma(u)$ are uniformly computable in ordinary time $O(s(n))$.
		\end{enumerate}
	\end{lemma}
	
	\begin{proof}
		By nondeterministic-space complementation, both $L$ and its complement
		have nondeterministic $O(s)$-space acceptors. Construct a machine $N$
		that first chooses which acceptor to run and uses at most two branches
		at each step. When the acceptor for $L$ accepts, $N$ halts with value
		$1$. When the acceptor for the complement accepts, $N$ halts with
		value $0$. Call these \emph{definite halts}. All other halts are
		\emph{inconclusive}. For each input $x$, some definite halt is
		reachable from the initial configuration $c_0$, and every such halt
		has value $v_x=\one_L(x)$.
		
		A configuration of $N$ can be encoded in $O(s)$ bits, including the
		control state, work-tape contents, and all head positions. For a fixed
		input, there are at most $2^{Cs}$ such configurations for some fixed
		positive integer $C$. A shortest path from $c_0$ to a definite halt
		has no repeated configuration. Give the message a counter of $Cs+1$
		bits and let $J_{\max}$ be its largest value. The counter can then
		record every step of this path without reaching $J_{\max}$.
		
		Write a message as $(c,\kappa,v)$, where $c$ is a purported
		configuration, $\kappa$ is the counter, and $v$ is a label bit. Set
		$\gamma(c,\kappa,v)=v$. Fixed field positions make the counter and
		label readable even when $c$ is malformed. For $j\in\{0,1\}$, define
		$f_j$ by applying the first applicable rule below.
		\begin{enumerate}
			\item If $c$ is malformed, $c$ is an inconclusive halt, or
			$\kappa=J_{\max}$, reset to $(c_0,0,v)$.
			\item If $c$ is a definite halt with value $w$, reset to $(c_0,0,w)$.
			\item Otherwise, take the successor indexed by $j$, increment $\kappa$,
			and preserve $v$. If branch $j$ is unavailable or its successor lies
		outside the encoded space, use a fixed inconclusive halt as the
		successor. Every configuration reachable from $c_0$ fits the encoding,
		so this convention never changes a genuine computation.
		\end{enumerate}
		
		From any message, every sequence of updates reaches a reset within
		$J_{\max}+1$ updates, since each update before the reset increments the
		counter. A reset leaves the message at $(c_0,0,v)$ for some bit $v$.
		Now choose the branches along a shortest path to a definite halt.
		The counter is still below $J_{\max}$ at that halt, so the next update
		uses the second rule and produces
		\[
		z_x=(c_0,0,v_x).
		\]
		This proves~(1), even if the initial message encoded an unreachable
		configuration with an incorrect halt value.
		
		Starting from $z_x$, the only rule that can change the label reads the
		value of a definite halt. Between resets, such a halt can be reached
		only by following genuine transitions from $c_0$, so its value is
		$v_x$. All other updates preserve the label. This proves~(2).
		
		For~(3), encode each work tape with a mark at its head position and
		encode the input-head position by a mark on a track of length $O(n+1)$.
		A fixed number of sequential scans suffices to check the encoding,
		read the scanned symbols, and write the successor configuration. In
		particular, scanning the input alongside its position track locates
		the scanned input symbol in $O(n+1)$ time. Incrementing the counter,
		writing a reset message, and reading the label each take $O(s)$ time.
		Thus each map and the readout run in ordinary time $O(s)$, even when
		reaching $z_x$ requires exponentially many map applications.
	\end{proof}
	
	\subsection{Sequential writes and supported choices}
	
	The program stores messages in two register slots. A pointer identifies
	the slot that holds the current message. To prepare an update, the
	program writes the new message into the other slot and changes the
	pointer only after the write is complete. A transmission during the
	write therefore carries a register word that still decodes to the old
	message. Once the pointer changes, the word decodes to the new one.
	Lemma~\ref{lem:compiler} gives the construction.
	
	A single descent can include several backward departures. To keep a
	later departure from applying a second update, the program stores a
	\emph{phase bit} alongside the message. Phase $0$ means \emph{ready},
	and phase $1$ means \emph{done}. A \emph{window} is an interval of
	logical departure indices on which the same update rule applies. There
	is a choice window for each $j\in\{0,1\}$. An executed backward move
	from that window sends a ready message $u$ to $f_j(u)$ and marks it
	done. Both choice windows preserve a done message. A reset window
	below the choice windows restores the ready phase while preserving
	the message. We call the change in the decoded message alone, with
	the phase and storage details omitted, a \emph{projected transition}.
	
	\begin{lemma}[Supported choices and projected transitions]\label{lem:gating}
		Let the messages be as in Lemma~\ref{lem:compiler}, let $f_0,f_1$ be
		total maps on them, and let $\gamma$ be a bit-valued readout. Suppose all three are computable in ordinary
		time $O(s)$. There is a compiled program with readout $\gamma$ and a
		phase bit $e\in\{0,1\}$ with the following properties for every pair
		of supported jumps $+a,-b$ with $a,b>0$.
		\begin{enumerate}
			\item From every joint state, a finite supported path reaches a ready
			state at some clock index $i<b$.
			\item From any ready state at an index $i<b$ with message $u$, and for
			each $j\in\{0,1\}$, a finite supported path reaches a ready state
			at an index below $b$ with message $f_j(u)$.
			\item On every supported transition, if the current message is $u$,
			the next message belongs to $\{u,f_0(u),f_1(u)\}$.
		\end{enumerate}
		The program uses $O(s)$ ordinary time and a register of length $O(s)$.
	\end{lemma}
	
	The construction in \ref{app:compiler} uses three windows
	with starting indices $t_R<t_0<t_1$. Their spacing accounts for the
	largest supported jump. Table~\ref{tab:windows} gives the transmission
	rule for each departure index. \Cref{fig:compiler} shows a supported
	path through the windows.
	
	\begin{table}[htbp]
		\centering\small
		\caption{Decoded message and phase transmitted by an executed backward
			move when the current message is $u$.}
		\label{tab:windows}
		\begin{tabular}{@{}lll@{}}
			\toprule
			Departure index $i$ & Ready before the move ($e=0$)
			& Done before the move ($e=1$)\\
			\midrule
			$i<t_R$           & $(u,0)$      & $(u,1)$\\
			$t_R\leq i<t_0$   & $(u,0)$      & $(u,0)$\\
			$t_0\leq i<t_1$   & $(f_0(u),1)$ & $(u,1)$\\
			$i\geq t_1$       & $(f_1(u),1)$ & $(u,1)$\\
			\bottomrule
		\end{tabular}
	\end{table}
	
	The paths implementing the two choices may have very different
	probabilities. For definite equilibrium readout, it suffices that each
	required finite path has positive probability on the given input.
	
	\subsection{Lifting the maps to a stationary joint process}
	
	The choice windows let the joint process follow a prescribed sequence
	of message updates. We use this freedom to reach the distinguished
	message $z_x$ from any recurrent state. Recurrence then provides a path
	back to the starting state, and every message along that return path
	has the correct label.
	
	\begin{proposition}\label{prop:lower}
		Under~\eqref{eq:clock-assumption}, $\NSPACE(s)\subseteq\Eq_\mu[O(s)]$.
	\end{proposition}
	
	\begin{proof}
		Fix $L\in\NSPACE(s)$. Choose a supported positive jump $+a$, and let
		$b$ be the least positive integer for which $-b$ is supported. Apply
		Lemma~\ref{lem:gating} to the maps $f_0,f_1$ and readout $\gamma$ of
		Lemma~\ref{lem:maps}. Every register word decodes to a message $u$ and
		a phase, and the compiled program returns $\gamma(u)$ as its final
		answer. Its register length and ordinary running time are both $O(s)$.
		
		Fix an input $x$. By Lemma~\ref{lem:gating}(1), every joint state can
		reach a ready boundary state. Let $u$ be its decoded message.
		Lemma~\ref{lem:maps}(1) gives a finite sequence of updates taking $u$
		to $z_x$. By Lemma~\ref{lem:gating}(2), each update in this sequence
		can be realized by a supported path ending at another ready boundary
		state. Concatenating these paths gives a supported path to a state
		whose decoded message is $z_x$.
		
		Now let $\omega$ be any recurrent joint state. Its communicating class
		is closed, so the path from $\omega$ to a state $w$ with message $z_x$
		stays in that class. Hence $w\leadsto\omega$. Along this return path,
		each transition either preserves the decoded message or applies one of
		$f_0,f_1$, by Lemma~\ref{lem:gating}(3). The message at $\omega$ is
		therefore reachable from $z_x$ by a finite sequence of the two maps.
		Lemma~\ref{lem:maps}(2) gives its label as $\one_L(x)$.
		
		Thus the compiled program has the correct final bit at every recurrent
		joint state. By Lemma~\ref{lem:tight}, a stationary joint law exists
		and every stationary joint law gives probability one to that bit.
		The program therefore decides $L$ by definite equilibrium readout.
	\end{proof}
	
	\subsection{Disabling feedback and fixing the resource scale}
	
	When feedback is disabled, each register word stays fixed at every
	clock tick. Each word can therefore be tested separately by a
	stationary law. To obtain a strict separation at the same resource
	scale, we use the deterministic space hierarchy. A bound $f(n)$ is
	\emph{space constructible} if a deterministic machine can mark out
	$f(n)$ work-tape cells on inputs of length $n$ using $O(f(n))$ space.
	
	\begin{proposition}\label{prop:off}
		Under~\eqref{eq:clock-assumption}, for every fixed positive integer
		$\alpha$ and $s(n)=(n+1)^\alpha$,
		\[
		\Eq_\mu^{\rm off}[O(s)]=\DTIME(s)\subsetneq\NSPACE(s).
		\]
	\end{proposition}
	
	\begin{proof}
		Let $\varpi$ be the clock marginal of a stationary joint law supplied
		by Lemma~\ref{lem:tight}. Since the clock is autonomous, $\varpi$ is
		stationary for the clock alone. Fix an input $x$ and a register word
		$r\in R_n$. With feedback disabled, the product law
		$\varpi\otimes\delta_r$, where $\delta_r$ is the point mass at $r$,
		is stationary: one tick preserves the clock law $\varpi$ and leaves
		the register at $r$. Definition~\ref{def:eq} applied to this law gives
		$\out_x(r)=\one_L(x)$.
		
		In particular, an ordinary deterministic machine can prepare the
		all-zero register word $0^{\ell(n)}$ and run the program on it. Since
		$\ell(n)=c(n+1)^\alpha$ with $c$ and $\alpha$ fixed, preparing the
		register takes $O(s)$ time, as does the run itself. This gives
		$\Eq_\mu^{\rm off}[O(s)]\subseteq\DTIME(s)$. Conversely, a
		deterministic $O(s)$-time decider can ignore the register. Its final
		bit is then correct for every register word, so it decides the same
		language by equilibrium readout with feedback disabled.
		
		For the strict inclusion, the time--space simulation of Hopcroft,
		Paul, and Valiant~\citep{HopcroftPaulValiant1977} gives
		\[
		\DTIME(s)\subseteq\DSPACE(s/\log s)
		\]
		for the fixed polynomial time bound $s(n)=(n+1)^\alpha$. This $s$ is
		space constructible, and $s/\log s=o(s)$, including when $\alpha=1$.
		The deterministic space hierarchy~\citep[Theorem~4.8]{AroraBarak2009}
		therefore gives a language in $\DSPACE(s)$ outside
		$\DSPACE(s/\log s)$. Hence
		\[
		\DTIME(s)\subseteq\DSPACE(s/\log s)
		\subsetneq\DSPACE(s)\subseteq\NSPACE(s).
		\]
		Together with Propositions~\ref{prop:upper} and~\ref{prop:lower},
		this proves Theorem~\ref{thm:main}.
	\end{proof}
	
	\begin{remark}[Classical consequences]\label{rem:consequences}
		At the linear scale, Theorem~\ref{thm:main} gives
		$\Eq_\mu[O(n+1)]=\NSPACE(n)$. By Kuroda's theorem, this is the class
		of context-sensitive languages, up to the convention for the empty
		word: a context-sensitive grammar cannot generate it, while a
		space-bounded machine may accept it~\citep{Kuroda1964}.
		
		At each fixed scale $s(n)=(n+1)^\alpha$, the class $\Eq_\mu[O(s)]$
		is closed under union and intersection. It is also closed under
		complement by the nondeterministic-space complementation
		theorem~\citep{Immerman1988,Szelepcsenyi1988}. The nondeterministic
		space hierarchy gives a strict increase in computational power with
		each increase in the integer exponent $\alpha$~\citep{Ibarra1972,Seiferas1977}.
		Savitch's theorem gives the deterministic upper bound
		$\Eq_\mu[O(s)]\subseteq\DSPACE(s^2)$~\citep{Savitch1970}.
		
		Proposition~\ref{prop:off} separates computation with and without
		feedback at every fixed polynomial scale. Taking the union over all
		positive integer exponents gives $\mathsf{PSPACE}$ with feedback, by
		Savitch's theorem, and $\mathsf{P}$ with feedback disabled. Whether
		feedback increases computational power when the exponent may vary
		from program to program is therefore exactly the open question
		$\mathsf{P}\ne\mathsf{PSPACE}$.
	\end{remark}
	
	\section{Observation from a blank start}\label{sec:observation}
	
	Equilibrium readout samples a joint state $(i,r)$ from a stationary
	law and returns the final bit $\out_x(r)$ of the selected history.
	We now start the same process at $(0,0^{\ell(n)})$. At tick $t$, we
	inspect the control state at index $X_t$ of the logical history with
	initial register word $R_t$.
	
	Negative drift supplied the stationary laws used in the classification.
	For the clock started at zero, Lemma~\ref{lem:tight} also bounds the
	probability of a large index at any given tick, uniformly in $t$.
	Because the program reads its input sequentially, reaching a distant
	input cell requires a large logical index. We use this connection to
	bound what can be observed from a blank start.
	
	\subsection{The initialized answer rule}
	
	On input $x$ of length $n$, start the joint process at
	$(0,0^{\ell(n)})$, where $\ell(n)=c(n+1)^\delta$ for fixed positive
	integers $c$ and $\delta$. In particular, $\ell(n)\geq n+1$.
	Each selected logical history begins with blank work tapes and the
	initial head positions specified in Section~\ref{sec:model}. At tick
	zero and after every subsequent tick $t$, inspect the control state
	of $C_{X_t}^{x,R_t}$. Define the first accepting tick by
	\[
	\tau_x=\inf\{t\in\N :
	C_{X_t}^{x,R_t}\text{ has an accepting control state}\}.
	\]
	Thus $\tau_x=0$ if the initial observation accepts, and $\tau_x=+\infty$
	if no observation ever accepts. We write $\Prob_{(i,r)}$ for probability
	when the joint process starts at $(i,r)$. The observation upper bound
	also applies to ordinary programs that run forever.
	
	A \emph{tick budget} is a nondecreasing function $Q\colon\N\to\N$
	satisfying $Q(n)\geq n+1$ for every $n$. We call $Q$
	\emph{subexponential} if $\log Q(n)=o(n)$. Equivalently, for every
	$\varepsilon>0$, we have $Q(n)\leq 2^{\varepsilon n}$ for all
	sufficiently large $n$.
	
	\begin{definition}[Observation classes]\label{def:observation}
		Fix a proposal law $\mu$ and a tick budget $Q$. A language $L$ belongs
		to $\mathsf{Obs}_\mu^{\rm init}[Q]$ if there are a uniform program
		with the register length and sequential access specified above and a
		constant $\beta>0$ such that the following conditions hold for every
		input $x$ of length $n$.
		
		For each member $x\in L$, the program is observed accepting by tick
		$Q(n)$ with probability at least $\beta$:
		\[
		\Prob_{(0,0^{\ell(n)})}\{\tau_x\leq Q(n)\}\geq\beta.
		\]
		For each nonmember, \emph{initialized soundness} requires
		\begin{equation}\label{eq:init-soundness}
			x\notin L \quad\Longrightarrow\quad
			\Prob_{(0,0^{\ell(n)})}\{\tau_x<+\infty\}=0.
		\end{equation}
		
		The class $\mathsf{Obs}_\mu[Q]$ uses the same acceptance condition
		for members and requires \emph{whole-register soundness} for
		nonmembers. For every $x\notin L$, every $r\in R_n$, and every
		$i\in\N$, the control state of $C_i^{x,r}$ must be nonaccepting.
		Write $\mathsf{Obs}^{\rm off}_\mu[Q]$ for this whole-register class
		with transmission disabled. The program may write within a history,
		and the initial register word of the observed history stays blank.
	\end{definition}
	
	For a fixed positive integer $\alpha$, set $s_\alpha(n)=(n+1)^\alpha$.
	In any observation class, $2^{O(s_\alpha)}$ denotes the union over
	budgets $Q(n)=2^{C s_\alpha(n)}$ for positive integers $C$. One
	constant $C$ serves all inputs of a given program. The register
	exponent $\delta$ may differ from the budget exponent $\alpha$.
	
	Condition~\eqref{eq:init-soundness} requires \emph{all-time initialized
		soundness}. The deadline $Q(n)$ applies only to members. For a
	nonmember, observing acceptance at any finite tick violates soundness,
	even after the deadline. With finite proposal support, this condition
	is equivalent to requiring that no finite sequence of supported
	proposals from the blank start yields an accepting observation. Indeed,
	each such sequence has positive probability, and every observation at
	a finite tick is determined by a finite sequence. The empty sequence
	covers the observation at tick zero.
	
	Initialized soundness constrains only observations reachable from the
	nonmember's blank start. Whole-register soundness covers every register
	word and every logical index, so for every tick budget $Q$,
	\[
	\mathsf{Obs}_\mu[Q]\subseteq\mathsf{Obs}_\mu^{\rm init}[Q].
	\]
	
	Table~\ref{tab:rules} compares the two answer rules. Theorems~A and~B
	describe what the same feedback process can decide under each rule.
	
	\begin{table}[htbp]
		\centering\small
		\caption{Equilibrium readout and observation from a blank start.
			Here $n=|x|$.}
		\label{tab:rules}
		\begin{tabularx}{\linewidth}{@{}l
				>{\raggedright\arraybackslash}X
				>{\raggedright\arraybackslash}X@{}}
			\toprule
			& Equilibrium readout (Definition~\ref{def:eq})
			& Initialized observation (Definition~\ref{def:observation})\\
			\midrule
			State sampled
			& $(i,r)$ drawn from a stationary joint law. At least one such law
			must exist.
			& $(X_t,R_t)$ at each tick $t\geq 0$, starting from
			$(0,0^{\ell(n)})$.\\
			What is read
			& The final bit $\out_x(r)$ of the history with initial register $r$.
			& The control state of $C_{X_t}^{x,R_t}$.\\
			Members
			& The bit is one almost surely under every stationary law.
			& Acceptance is observed by tick $Q(n)$ with probability at least
			$\beta$.\\
			Nonmembers
			& The bit is zero almost surely under every stationary law.
			& Initialized soundness rules out acceptance almost surely at all
			ticks. Whole-register soundness rules it out for every $(i,r)$.\\
			Halting
			& Required for every initial register word.
			& Optional.\\
			Resources
			& $O(s)$ ordinary steps per history and $O(s)$ register symbols.
			& $c(n+1)^\delta$ register symbols and a deadline of $Q(n)$ ticks
			for members.\\
			\bottomrule
		\end{tabularx}
	\end{table}
	
	Let $\RDEF$ denote the class of \emph{bounded-prefix languages}.
	A language $L$ belongs to $\RDEF$ if there is an integer $h_0\geq 0$
	such that any two words of length at least $h_0$ with the same first
	$h_0$ symbols either both belong to $L$ or both lie outside $L$.
	Membership of shorter words may be specified separately. These
	languages are also called \emph{reverse definite}~\citep[Section~4]{BrzozowskiLiLiu2012}.
	The corresponding condition for \emph{definite} languages uses a
	bounded suffix~\citep{PerlesRabinShamir1963}.
	
	\subsection{Budgets and clock height}
	
	To bound what the clock can observe by a given deadline, we need to
	control the greatest index it visits up to that tick.
	
	\begin{corollary}[Clock height bound]\label{cor:clockmax}
		Suppose $\mu$ satisfies~\eqref{eq:clock-assumption}, and let $\theta$
		and $A$ be the constants from Lemma~\ref{lem:tight}. For every tick
		count $Q\in\N$ and height $H\in\N$, the clock started at zero satisfies
		\begin{equation}\label{eq:clockmax}
			\Prob_0\!\left\{\max_{0\leq t\leq Q}X_t\geq H\right\}
			\leq (1+A)(Q+1)e^{-\theta H}.
		\end{equation}
	\end{corollary}
	
	\begin{proof}
		By~\eqref{eq:expbound} with $i=0$, we have
		$\E_0 e^{\theta X_t}\leq 1+A$ for every tick $t$. Applying Markov's
		inequality to $e^{\theta X_t}$ gives
		\[
		\Prob_0\{X_t\geq H\}
		=\Prob_0\{e^{\theta X_t}\geq e^{\theta H}\}
		\leq (1+A)e^{-\theta H}.
		\]
		The clock reaches height at least $H$ by tick $Q$ if this event occurs
		at any of the $Q+1$ ticks $t=0,\ldots,Q$. The union bound
		gives~\eqref{eq:clockmax}.
	\end{proof}
	
	\begin{theorem}[Observation from a blank start]\label{thm:observation}
		Suppose $\mu$ satisfies~\eqref{eq:clock-assumption}.
		\begin{enumerate}
			\item For every subexponential tick budget $Q$, that is, every tick
			budget satisfying $\log Q(n)=o(n)$,
			\[
			\mathsf{Obs}_\mu[Q]
			=\mathsf{Obs}_\mu^{\rm init}[Q]
			=\RDEF.
			\]
			\item For every fixed integer $\alpha\geq1$, let $s=s_\alpha$.
			Under whole-register soundness,
			\[
			\mathsf{Obs}_\mu[2^{O(s)}]=\NTIME(s),\qquad
			\mathsf{Obs}^{\rm off}_\mu[2^{O(s)}]=\DTIME(s).
			\]
			At $\alpha=1$, the class with transmission disabled is strictly
			contained in the class with feedback.
		\end{enumerate}
	\end{theorem}
	
	\begin{proof}[Proof of part~(1)]
		The inclusion from whole-register to initialized observation follows
		from Definition~\ref{def:observation}. We prove the two outer inclusions.
		
		Let $L\in\mathsf{Obs}_\mu^{\rm init}[Q]$, and fix an acceptance
		probability bound $0<\beta\leq 1$. With $\theta$ and $A$ from
		Lemma~\ref{lem:tight}, set
		\[
		H(n)=\left\lceil
		\frac{1}{\theta}
		\ln\!\left(\frac{4(1+A)(Q(n)+1)}{\beta}\right)
		\right\rceil.
		\]
		This height is nondecreasing because $Q$ is nondecreasing. Moreover,
		$H(n)=O(1+\log Q(n))=o(n)$.
		
		Fix a member $x$ of length $n$. From its blank start, the probability
		of observing acceptance by tick $Q(n)$ is at least $\beta$. The clock
		starts at zero, so Corollary~\ref{cor:clockmax} and the choice of
		$H(n)$ give
		\[
		\Prob_0\!\left\{\max_{0\leq t\leq Q(n)}X_t\geq H(n)\right\}
		\leq \beta/4.
		\]
		Thus, with probability at least $3\beta/4$, acceptance is observed by
		tick $Q(n)$ and the clock stays below $H(n)$ throughout those ticks.
		In particular, some finite supported proposal word observes acceptance
		from the blank start while staying below $H(n)$.
		
		Every member therefore has an accepting proposal word of height at
		most $H(|x|)$. Since $H$ is nondecreasing and sublinear, and since
		the program has all-time initialized soundness,
		Lemma~\ref{lem:sublinear-accepting-height} gives $L\in\RDEF$.
		Its proof uses Lemma~\ref{lem:register-transfer} to carry prefix
		agreement through changes of logical history. Both lemmas are proved
		in \ref{app:observation}.
		
		Conversely, let $L\in\RDEF$ and choose a prefix bound $h_0\geq 1$,
		enlarging it to one if necessary. A decider can ignore the register
		and store the prefix read so far in its finite control. It decides as
		soon as it reads either the $h_0$th symbol or the endmarker of a
		shorter word. In either case, the transition that reads the decisive
		symbol can enter the appropriate halting state. Thus its halting
		index satisfies
		\[
		T(x)\leq \min\{h_0,|x|+1\}.
		\]
		Only the accepting halt is an accepting control state, so the decider
		satisfies whole-register soundness.
		
		Choose a supported positive jump $+a$ and put
		$m(x)=\lceil T(x)/a\rceil$. From clock index zero, $m(x)$ consecutive
		proposals of $+a$ take the clock to index $m(x)a\geq T(x)$. The halted
		configuration persists at that index, so this proposal word observes
		acceptance whenever $x\in L$. Since $a\geq 1$, the word takes at most
		$|x|+1\leq Q(|x|)$ ticks. It also takes at most $h_0$ ticks, so its
		probability is at least $\mu(a)^{h_0}>0$, uniformly in $x$. This gives
		$L\in\mathsf{Obs}_\mu[Q]$ and completes the proof.
	\end{proof}
	
	In part~(1), the lower bound $Q(n)\geq n+1$ is used only to prove
	$\RDEF\subseteq\mathsf{Obs}_\mu[Q]$. It gives the bounded-prefix
	decider enough ticks to reach its halt even on short members. For the
	upper bound, it suffices that $Q$ be nondecreasing and that
	$\log(Q(n)+1)=o(n)$. The same choice of $H(n)$ then gives
	\[
	\mathsf{Obs}_\mu[Q]\subseteq\mathsf{Obs}_\mu^{\rm init}[Q]
	\subseteq\RDEF.
	\]
	
	\subsection{Exponential budgets and certificate search}
	
	\begin{proof}[Proof of part~(2)]
		Fix $\alpha\geq1$, abbreviate $s=s_\alpha(n)$, and take a program
		recognizing $L$ under whole-register soundness by tick $Q_C(n)=2^{Cs}$
		with success probability at least $\beta>0$. Choose a positive
		integer $D$ large enough that for $H=Ds$,
		\[
		(1+A)(Q_C(n)+1)e^{-\theta H}<\beta/2
		\qquad\text{for every }n.
		\]
		Such a constant exists because $\log(Q_C(n)+1)=O(s)$.
		The simulator can use this fixed integer $D$ without evaluating the
		clock probabilities.
		By Corollary~\ref{cor:clockmax}, each member has positive probability
		of being observed accepting by the deadline while the clock stays
		below $H$. On such a path, some history with initial register word
		$r$ has an accepting control state before ordinary step $H$.
		
		A nondeterministic multitape machine guesses the first
		$\min\{\ell(n),H\}$ symbols of $r$, then simulates the ordinary
		program from that prefix for up to $H$ steps. It checks the control
		state at index zero and after each step. No transition in the
		simulation can read beyond cell $H$. If $\ell(n)<H$, the simulator
		places the register endmarker in its prescribed position. Every
		guessed prefix thus extends to a genuine register word, and
		whole-register soundness rules out an accepting guess on a nonmember.
		
		The simulator can prepare this prefix and its time bound in $O(s)$
		steps even if the register and budget exponents differ. After scanning
		the input, it forms a unary block of length $H=D(n+1)^\alpha$ by a
		fixed number of nested scans, as in Lemma~\ref{lem:compiler}. If
		$\delta\leq\alpha$, the same construction marks the full register
		length $c(n+1)^\delta$ in time $O(s)$, so the simulator can stop at
		the earlier endpoint. If $\delta>\alpha$, then $\ell(n)\geq H$ for
		all but finitely many input lengths. The exceptions and their lengths
		are fixed by $c,\delta,D,\alpha$ and can be encoded in finite control.
		The machine returns its heads to their initial positions before
		simulating the run. This gives the $\NTIME(s)$ upper bound. With
		transmission disabled the initial word is always blank, so filling
		the same prefix with zeros gives a deterministic $O(s)$-time test.
		
		For the reverse inclusion with feedback, let a nondeterministic machine
		$N$ decide $L$ in time $O(s)$. Use binary branching, at constant
		overhead, and choose a positive integer $d$ so that all branches
		halt within $m=ds$ steps. A word $u$ of $m$ bits specifies one
		branch. A deterministic verifier $V(x,u)$ simulates $N$, reading
		at most one bit per step to choose a successor and ignoring unused
		bits after a halt. It rejects an invalid choice. The verifier takes
		$O(s)$ steps, and some $u$ gives $V(x,u)=1$ exactly when $x\in L$.
		
		Interpret $u$ as an integer written in $m$ bits, and set
		$g(u)=u+1\bmod 2^m$. Apply Lemma~\ref{lem:compiler} at scale $s$
		with this single map and readout $V(x,u)$. Two $m$-cell slots and a
		pointer fit in a register of length $cs$ for a suitable constant
		$c$, so the construction uses the allowed exponent $\delta=\alpha$.
		Ignore surplus cells and decode each slot symbol as a bit by a
		fixed rule that preserves zero and one. Let $h=O(s)$ be the index
		at which the pointer switches to the completed successor. Keep the
		verdict on the original $u$ in finite control. The transition that
		switches the pointer also enters the halt for that verdict, and only
		the accepting halt is accepting. At every index at least $h$ the
		observer therefore sees the verdict on $u$, while an executed
		backward move from such an index transmits $g(u)$. Earlier backward
		moves preserve the decoded word. On a nonmember the verifier
		rejects every word, so all histories satisfy whole-register soundness.
		
		At a tick $t$ with $X_t\geq h$, the observer sees the verdict on the
		current candidate. If the next move is an executed backward move, the
		register of the new history at tick $t+1$ contains the successor
		candidate. The enumeration counts verdicts seen at these departure
		ticks; the arrival index may lie below $h$. Each such move advances the
		certificate by exactly one, and all other ticks preserve it. The
		first $2^m$ such departures therefore show the verdict on every
		candidate, whatever the starting candidate. To bound the wait,
		choose supported jumps $+a,-b$ and arrange $h\geq b$. Put
		$k=\lceil h/a\rceil$. From any index, a block of $k$ proposals
		$+a$ followed by $-b$ guarantees one such departure. In disjoint
		blocks of length $k+1$ this event occurs independently with
		probability $\pi=\mu(a)^k\mu(-b)$. The mean wait in blocks for
		one success is $1/\pi$, so the expected number of ticks to finish
		$2^m$ successful blocks is
		\[
		(k+1)2^m/\pi = 2^{O(s)},
		\]
		since $m=ds$, $k=O(s)$ and the two proposal probabilities are fixed.
		Extra departures between successful blocks also advance the same
		enumeration. By Markov's inequality, twice this expected time gives
		success probability at least $1/2$ on every member. One integer
		$C$ makes $2^{Cs}$ at least this large for all $n$.
		
		Finally, a deterministic $O(s)$-time decider can ignore the register
		and make its accepting halt its only accepting state. If it halts by
		step $ds$, a block of $j=\lceil ds/a\rceil$ consecutive $+a$ proposals
		reveals the verdict from any starting index. In disjoint blocks the
		expected wait for the first such block is $j/\mu(a)^j=2^{O(s)}$
		ticks. Doubling the expected wait gives success probability at least
		$1/2$, proving the reverse inclusion with transmission disabled.
		At $\alpha=1$, strict inclusion follows from the multitape
		linear-time separation~\citep{PaulPippengerSzemerediTrotter1983}.
	\end{proof}
	
	For example, let $\mathrm{PARITY}$ be the set of binary words with
	an odd number of ones. A single scan decides $\mathrm{PARITY}$ in
	linear time, so part~(2) at $\alpha=1$ gives a $2^{O(n)}$ tick budget.
	For every $h\geq 0$, the words $0^h$ and $0^h1$ agree on their first
	$h$ symbols and have opposite parity. Thus $\mathrm{PARITY}$ lies
	outside $\RDEF$, and part~(1) rules out every subexponential tick
	budget even under initialized soundness.

	For any fixed integer $\alpha\geq2$, strictness in part~(2) is the
	open question whether deterministic and nondeterministic multitape time
	differ at the bound $(n+1)^\alpha$. Santhanam extended the
	linear-time separation to time $n\sqrt{\log^* n}$~\citep{Santhanam2001},
	a bound that grows more slowly than quadratic time.
	
	\begin{corollary}[Observation over all polynomial scales]\label{cor:obs-polynomial}
		For every clock law satisfying~\eqref{eq:clock-assumption}, under
		whole-register soundness,
		\[
		\bigcup_{\alpha\geq1}\mathsf{Obs}_\mu[2^{O(s_\alpha)}]=\NPclass,
		\qquad
		\bigcup_{\alpha\geq1}\mathsf{Obs}^{\rm off}_\mu[2^{O(s_\alpha)}]
		=\Pclass .
		\]
		The unions are over positive integer exponents. They differ if and
		only if $\Pclass\ne\NPclass$.
	\end{corollary}
	
	\begin{proof}
		Take the union of the equalities in Theorem~\ref{thm:observation}(2).
		Every polynomial time bound is $O(s_\alpha)$ for some $\alpha$, and
		each $s_\alpha$ is polynomial. The two unions of time classes are
		therefore $\NPclass$ and $\Pclass$, respectively.
	\end{proof}

	Under initialized soundness alone, exponential budgets reach polynomial
	space over all scales.
	
	\begin{proposition}[Exponential budgets under initialized soundness]\label{prop:init-pspace}
		Suppose $\mu$ satisfies~\eqref{eq:clock-assumption}, and fix a
		polynomial scale $s=s_\alpha$. Then
		\[
		\DSPACE(s)\subseteq\mathsf{Obs}^{\rm init}_\mu[2^{O(s)}]
		\subseteq\NSPACE(s).
		\]
		Consequently,
		$\bigcup_{\alpha\geq1}\mathsf{Obs}^{\rm init}_\mu[2^{O(s_\alpha)}]=\PSPACE$.
	\end{proposition}
	
	\begin{proof}
		\emph{Upper bound.} Fix a program, an integer $C$ and a constant
		$\beta>0$ that witness $L\in\mathsf{Obs}^{\rm init}_\mu[2^{Cs}]$, and
		put $Q=2^{Cs(n)}$. Let $\theta$ and $A$ be the constants of
		Corollary~\ref{cor:clockmax}, and choose an integer $D$ with
		$2(1+A)e^{C\ln 2-\theta D}<\beta/2$. Since $s(n)\geq1$, the height
		$H=Ds(n)$ satisfies $(1+A)(Q+1)e^{-\theta H}<\beta/2$ for every $n$.
		By Corollary~\ref{cor:clockmax}, on every member some finite
		supported path of at most $Q$ ticks from the blank start shows
		acceptance while the clock index stays below $H$.
		
		A nondeterministic machine searches for such a path. It stores the
		tick count, the current index $i<H$, and the first
		$\min\{\ell(n),H\}$ cells of the initial register word of the
		current history, starting from index zero and the blank prefix. At
		each tick it reconstructs the current configuration by running the
		program for $i$ steps from the stored prefix with blank work tapes,
		and it accepts if the control state accepts. Otherwise it rejects
		at tick $Q$, or it guesses a supported proposal and rejects the
		branch if the new index is at least $H$. On an executed backward
		move it replaces the stored prefix by the register prefix of the
		reconstructed configuration at the departure index. Forward moves
		and holds keep the stored prefix.
		
		The reconstruction is exact. A run of fewer than $H$ steps neither
		reads nor writes a register cell beyond the first $H$, so along the
		path every register cell beyond the first $H$ stays blank. When
		$\ell(n)\leq H$, the machine also supplies the right endmarker of
		the register. By induction on the ticks, the stored index and prefix
		agree with those of the actual path. The tick count uses $Cs(n)+1$
		bits, and the prefix and the simulated tapes use $O(H)=O(s)$ cells.
		A member therefore has an accepting branch. An accepting branch on a
		nonmember would be a finite supported path from the blank start that
		shows acceptance. Such a path has positive probability, so
		initialized soundness excludes it. Hence $L\in\NSPACE(s)$.
		
		\emph{Lower bound.} Let a deterministic machine $A$ decide $L$ in
		space $O(s)$ and halt on every input. Encode the configurations of
		$A$ on inputs of length $n$ as binary words of length $m(n)=ds(n)$
		for a fixed integer $d$, with tracks for the tape symbols and head
		marks and a field for the control state. A track marking the
		input-head position fits because $n+2\leq 2s(n)$. Reserve the
		all-zero word for the start configuration and mark every other code
		with a tag bit. Let $g_x$ send the code of each configuration to the
		code of its successor and fix every halting configuration. It sends
		every word that encodes no configuration, and every configuration
		whose successor does not fit the encoding, to the code of a fixed
		rejecting halt. The run of $A$ on $x$ always fits. Let
		$v_x(u)=1$ exactly when $u$ encodes an accepting halting
		configuration. A constant number of sequential scans computes $g_x$
		and $v_x$ in ordinary time $O(s)$, and $m(n)$ cells can be marked
		out in $O(s)$ steps as in the proof of Lemma~\ref{lem:compiler}.
		
		Apply Lemma~\ref{lem:compiler} with the single map $g_x$ and readout
		$v_x$, with commit index $h=O(s)$ chosen so that $h\geq b$. The
		register has length $O(s)$, so the construction uses the exponent
		$\delta=\alpha$. Keep the saved verdict in finite control. The
		transition that switches the pointer also enters the halt for that
		verdict, and only the accepting halt is accepting. The blank register
		selects the all-zero slot, which encodes the start configuration. By
		induction from the blank start, every decoded message encodes a
		configuration on the run of $A$ on $x$, and each tick keeps it or
		advances it by one step. A backward departure from an index below
		$h$ transmits the same message, and one from an index at least $h$
		transmits its successor. An accepting control state appears only at
		an index at least $h$ of a history whose message encodes the
		accepting halt of $A$. A nonmember is therefore never observed
		accepting.
		
		On a member, $A$ cannot repeat a configuration before it halts, so
		it halts within $N=2^{es(n)}$ steps for a fixed integer $e$. As in
		the proof of Theorem~\ref{thm:observation}(2), put $k=\lceil
		h/a\rceil$. From any index, a block of $k$ proposals $+a$ followed by
		$-b$ produces an executed backward departure from an index at least
		$h$. Disjoint blocks succeed independently with probability
		$\pi=\mu(a)^k\mu(-b)$. After $N$ successful blocks the message
		encodes the accepting halt, which $g_x$ fixes. The departure tick of
		one more successful block shows the accepting verdict. The expected
		number of ticks to complete $N+1$ successful blocks is
		\[
		(k+1)(N+1)/\pi=2^{O(s)},
		\]
		since $k=O(s)$ and the two proposal probabilities are fixed. By
		Markov's inequality, twice this expected time gives success
		probability at least $1/2$, and one integer $C$ makes $2^{Cs}$ at
		least this large for all $n$.
		
		\emph{Union over scales.} Every polynomial space bound is
		$O(s_\alpha)$ for some $\alpha$. Conversely, Savitch's theorem gives
		$\NSPACE(s_\alpha)\subseteq\DSPACE(s_\alpha^2)$~\citep{Savitch1970},
		and $s_\alpha^2=s_{2\alpha}$ is polynomial.
	\end{proof}
	
	\begin{remark}[Charging the reconstruction work]\label{rem:work}
		Charge each tick one step, plus the ordinary steps needed to
		reconstruct the configuration in view, plus $\ell(n)$ steps for
		copying the register when a backward move is executed. At index $i$
		the reconstruction takes at most $i$ steps.
		Theorem~\ref{thm:observation} and Proposition~\ref{prop:init-pspace}
		remain true when the budget bounds this total charge instead of the
		number of ticks.
		
		Each tick costs at least one step, so a bound on the total charge
		also bounds the number of ticks. For the other direction, let a
		program and a constant $\beta>0$ witness membership with tick budget
		$Q$. With $\theta$ and $A$ as in Corollary~\ref{cor:clockmax},
		choose $H=O(\log Q(n))$ such that
		$(1+A)(Q(n)+1)e^{-\theta H}\leq\beta/2$. By that corollary, with
		probability at least $\beta/2$ the program is observed accepting by
		tick $Q(n)$ and the index stays below $H$ until then. On this event
		the total charge is at most $(Q(n)+1)(H+\ell(n)+1)$. Since $\ell$
		is polynomial and $Q(n)\geq n+1$, this bound is $2^{O(s)}$ when
		$Q=2^{O(s)}$, and its logarithm is $o(n)$ when $\log Q(n)=o(n)$.
		The two soundness rules do not mention time, so they are not
		affected.
	\end{remark}
	
	We adapt three arguments from our study of random clocks without
	feedback~\citep{DidehvarFarhat2026RandomClocks}. They are the clock
	height bound in Corollary~\ref{cor:clockmax} (Theorem~4.1 and Lemma~4.2
	there), the shortening argument in
	Lemma~\ref{lem:sublinear-accepting-height} (Lemmas~2.13, 2.14
	and~2.17 there), and the use of consecutive upward proposals in the
	proof of part~(2) (Lemma~2.11 there). We include the
	arguments to keep the proofs self-contained.
	
	The new step in the shortening argument is
	Lemma~\ref{lem:register-transfer}, which preserves prefix agreement
	through any number of history replacements. All-time initialized
	soundness matters here because an accepting path transferred to a
	shorter input may take more ticks than the shorter input's budget
	allows. If the shorter input were a nonmember, that path would violate
	all-time soundness. Soundness only through the deadline would allow
	such a path, leaving that version of the observation problem open here.
	
\section{Operational costs}\label{sec:costs}

The certificate search of Section~\ref{sec:observation} advances each
time a backward move leaves an index where the next candidate is ready
to send. Here we fix one index at which updates become available and
measure the wait for successive updates under a nearest-neighbor clock.
The latch of Section~\ref{sec:latch} needs one such update. A counter
that fills at $M$ needs $M$, and erasures can force it to start over.

\subsection{Useful updates and waiting times}

Let $(K_t)_{t\in\N}$ be independent proposals, each equal to $+1$ with
probability $p$ and $-1$ with probability $q=1-p$, where
$0<p<\tfrac12$. At tick $t$, the clock starts at $X_t$ and applies
$K_t$ to reach $X_{t+1}=\max\{0,X_t+K_t\}$. Thus a downward proposal
at index zero is held. Put $\rho=p/q$. The clock has the stationary law
$\varpi(i)=(1-\rho)\rho^i$ for $i\in\N$. The detailed balance equations
$\varpi(i)p=\varpi(i+1)q$ hold for each neighboring pair, and the
probabilities sum to one. The clock is irreducible, so this law is
unique.

Fix a positive integer $h$, called the \emph{departure height}. A
\emph{useful update} occurs if the clock proposes $-1$ from an index
$X_t\geq h$. The move is executed because $h\geq1$, and the condition
uses the index before the move. A one-map implementation from
Lemma~\ref{lem:compiler} has $h$ equal to its commit index, the first
index at which the updated message is ready to send. Such an
implementation applies its map at every useful update and preserves
the interpreted message on earlier departures.

Write $I_t=\one\{X_t\geq h,\ K_t=-1\}$ for the indicator of a useful
update on tick $t$. Let $U_t$ count useful updates among the first
$t$ ticks. For a positive integer $M$, let $\tau_M$ be the number of
ticks needed to see $M$ updates. Thus
\[
 U_t=\sum_{j=0}^{t-1}I_j,\qquad
 \tau_M=\inf\{t\in\N:U_t\geq M\}.
\]
Here $U_0=0$. Write $\Prob_0$ and $\E_0$ when the clock starts at
zero. For a target probability $0<\beta<1$, the least tick budget
that gives $M$ useful updates with probability at least $\beta$ is
\[
 t_\beta(p;h,M)=\inf\{t\in\N:\Prob_0(\tau_M\leq t)\geq\beta\}.
\]
Under the stationary clock law, the probability of being at or above
$h$ is $\rho^h$. The next proposal is independent of the index, so
each tick gives a useful update with probability $q\rho^h$. From zero,
the bound $\E_0 U_t\leq qt\rho^h$ in \ref{sec:costproofs} gives the
lower bound on the tick budget.

\begin{theorem}[Matching update cost]\label{thm:waiting}
Put $C(p)=q/(q-p)^2+1/q$. For positive integers $h$ and $M$, an integer
$t\geq0$, and $0<\beta<1$,
\begin{align}
 \Prob_0(\tau_M\leq t)&\leq\frac{q\,t\,\rho^h}{M},\label{eq:waitinglower}\\
 \E_0\tau_M&\leq C(p)M\rho^{-h},\label{eq:waitingmean}\\
 \frac{\beta}{q}M\rho^{-h}
 &\leq t_\beta(p;h,M)
 \leq\ceil{\frac{C(p)}{1-\beta}M\rho^{-h}}.\label{eq:waitingquantile}
\end{align}
In particular, $t_\beta(p;h,M)=\Theta_{p,\beta}(M\rho^{-h})$.
\end{theorem}

\subsection{A counter with whole-message erasures}

Let $Y_t\in\{0,\ldots,M\}$ be the value of the stored counter at the
start of tick $t$. The counter is \emph{saturating}: a useful update
adds one up to the maximum $M$. Let $E_t$ be the indicator of an
erasure on tick $t$, with $\Prob(E_t=1)=\varepsilon\in[0,1]$. The
erasure draws are independent of one another, the initial state, and
the proposals. An erasure clears the entire counter
and overrides the update even when $I_t=1$. Specifically,
\[
 Y_{t+1}=
 \begin{cases}
  0, & E_t=1,\\
  \min\{Y_t+I_t,M\}, & E_t=0 .
 \end{cases}
\]
The clock always applies $K_t$ to update $X_t$. The pair $(X_t,Y_t)$
is a Markov chain. A stationary law is a joint distribution of the
pair left unchanged by one tick, and the clock position and counter
can be correlated under that law. We write $\Prob(Y=M)$ for the
probability of finding the counter full under the stationary law
being considered.

\begin{theorem}[Erasure threshold for the counter]\label{thm:erasure}
For every $h,M\geq1$ the clock-counter chain has stationary probability laws. For any such law, if $\varepsilon>0$,
\begin{equation}\label{eq:erasurenec}
 \Prob(Y=M)\leq\frac{(1-\varepsilon)q\rho^h}{\varepsilon M}.
\end{equation}
For $0<\beta<1$, put $D(p,\beta)=2C(p)/(1-\beta)+1$ and
$c(p,\beta)=(1-\beta)/(2D(p,\beta))$.
If $0\leq\varepsilon\leq c(p,\beta)\rho^h/M$, then every stationary
law satisfies $\Prob(Y=M)\geq\beta$. Conversely, $\Prob(Y=M)\geq\beta$
requires
\begin{equation}\label{eq:necessaryeps}
 \varepsilon\leq\frac{q\rho^h}{\beta M+q\rho^h}
 \leq\frac{q}{\beta}\frac{\rho^h}{M}.
\end{equation}
The necessary and sufficient bounds have the same dependence on $h$
and $M$, up to constants depending only on $p$ and $\beta$.
\end{theorem}

For an $m$-bit counter that uses all its values, $M=2^m-1$. The
waiting time grows with the number of increments as well as with the
height at which each increment becomes ready. Preserving progress
across handoffs is a concern shared with checkpoint and restart
methods~\citep[Section~1]{ElnozahyEtAl2002}. The erasure rule also gives
an instance of stochastic resetting~\citep{EvansMajumdar2011} in which
the message resets while the clock continues.

\subsection{A numerical latch example}\label{sec:numerical-example}

Take $p=1/3$, $q=2/3$, $\rho=1/2$, $M=1$ and $\beta=1/2$.
Here $C(p)=6+3/2=15/2$, so Theorem~\ref{thm:waiting} bounds the
median update budget between $(3/4)2^h$ and $15\cdot2^h$.
Equation~\eqref{eq:hittingexact} gives the mean time to first reach
height $h$ as $6(2^h-1)-3h$. From that first visit, the clock stays at
or above $h$ until the next downward proposal. The wait up to and
including that proposal is geometric with mean $1/q=3/2$. The mean time
to the first useful update is therefore $6(2^h-1)-3h+3/2$, which
gives $9/2$, $27/2$ and $69/2$ ticks at heights one, two and three.

For erasures, $D(p,\beta)=31$ and $c(p,\beta)=1/124$, so
$\varepsilon\leq2^{-h}/124$ ensures that the counter is full with
probability at least one half under every stationary law. At $h=1$,
the sufficient bound is $\varepsilon\leq1/248$ and
Equation~\eqref{eq:necessaryeps} gives the necessary bound
$\varepsilon\leq2/5$.
The constants leave a wide gap. The bounds agree in their dependence
on the departure height and target size.

	\section{Conclusion and open questions}\label{sec:discussion}
	
	Backward feedback realizes nondeterministic space at equilibrium and
	nondeterministic time from a blank start under whole-register soundness
	and a tick budget exponential in the chosen scale. The two proofs give
	different reasons for this parallel. At equilibrium, every recurrent
	class must carry the right answer. The clock need only allow the
	finite sequence of moves that brings a simulated space computation
	to a correct label. From a blank start, the observer must actually
	see acceptance by a deadline. Feedback can meet that deadline by
	enumerating certificates, each of which one ordinary run can verify.
	
	With transmission disabled, both classifications give deterministic
	time at the same polynomial scale. Taking the union over scales,
	the equilibrium comparison is $\Pclass$ versus $\PSPACE$, while
	the observation comparison under whole-register soundness is
	$\Pclass$ versus $\NPclass$ (Corollary~\ref{cor:obs-polynomial}).
	Under initialized soundness, observation with feedback reaches
	$\PSPACE$ over all scales (Proposition~\ref{prop:init-pspace}), as
	equilibrium readout does. The budget matters even for
	parity. With subexponentially many ticks in the input length, both
	soundness rules give only bounded-prefix languages.

	The counter shows what it takes to keep progress once it can be
	passed on. Without losses, it eventually fills and stays full.
	When handoffs need a late step in the run, reaching a full count
	from a blank start can take a long time. If the count is sometimes
	cleared, those same waits make it harder to rebuild. A larger count
	takes longer to fill, so it needs longer stretches without losses.
	
	\paragraph{Other clocks}
	The clock studied here draws each proposal independently from the same
	fixed law. Other models of computation that look back allow the program
	or an adversary to influence the proposals.
	\otherclockssource{}
	When the program chooses the proposals, equilibrium readout with
	feedback characterizes $\DSPACE(s)$; with adversary-chosen proposals,
	the class depends on the readout rule.
	
	\paragraph{Open questions}
	\begin{enumerate}
		\item \emph{Bounded-error equilibrium readout.}
		Fix a clock law satisfying~\eqref{eq:clock-assumption} with rational
		proposal probabilities. Which languages can be decided at scale $s$
		if every stationary law must give the correct final bit with
		probability at least $2/3$? For definite readout, a single recurrent
		state with the wrong bit witnesses failure. For bounded-error
		readout, the relevant quantity is the total stationary mass of such
		states under each law.
		
		\item \emph{Exponential budgets under initialized soundness.}
		For a fixed polynomial scale $s=s_\alpha$,
		Proposition~\ref{prop:init-pspace} places
		$\mathsf{Obs}^{\rm init}_\mu[2^{O(s)}]$ between $\DSPACE(s)$ and
		$\NSPACE(s)$. Which class is it? At linear scale the two bounds are
		deterministic and nondeterministic linear space, which are not known
		to differ.
		The verifier in Theorem~\ref{thm:observation}(2) may guess a
		register prefix that no history reachable from the blank start
		uses. Whole-register soundness makes that guess harmless on a
		nonmember. Initialized soundness leaves its verdict unconstrained.
		
		\item \emph{Soundness up to the observation deadline.}
		Which languages can an observer recognize from a blank start if a
		nonmember must have zero probability of acceptance only through tick
		$Q(|x|)$? In the prefix-transfer proof, an accepting path transferred
		to a shorter input can extend beyond that input's deadline. Such a
		path would be consistent with the weaker soundness condition.
		
		\item \emph{Index-dependent clocks.}
		Suppose the proposal law $\mu_i$ may depend on the current clock index
		$i$ and is independent of the program and register. Under what
		conditions on $(\mu_i)_{i\in\N}$ does the fixed-scale classification
		in Theorem~\ref{thm:main} continue to hold?
	\end{enumerate}
	\appendix
	\renewcommand{\thetheorem}{\Alph{section}.\arabic{theorem}}
	\counterwithin*{figure}{section}
	\counterwithin*{table}{section}
	\renewcommand{\theHfigure}{app.\Alph{section}.\arabic{figure}}
	\renewcommand{\theHtable}{app.\Alph{section}.\arabic{table}}
	
	\section{Sequential implementation of the feedback maps}
	\label{app:compiler}\label{sec:compiler}
	
	A backward move can transmit the register while the program is still
	writing an update. We therefore assign a decoded message $m(r)$ to
	every physical register word $r$, including malformed words and words
	transmitted during a write. The program prepares a complete word in
	an inactive slot. A \emph{commit} is the single ordinary transition
	that selects this slot for decoding.
	
	The \emph{compiler} takes a fixed finite list of message maps and
	constructs one sequential program that presents them in order. Each
	map is total, so it is defined on every decoded message. We suppress
	the input $x$ in the notation for the maps and the readout.
	
	\begin{lemma}[Atomic interpretation]\label{lem:compiler}
		For each input length $n$, let the messages be all binary words of
		length $k(n)=O(s(n))$, where a uniform procedure can mark out $k(n)$
		cells in $O(s(n))$ ordinary steps.
		Let $g_1,\ldots,g_J$ be a fixed finite list of total message maps,
		and let $\gamma$ be a bit-valued readout. Suppose each map and the
		readout can be computed from the input and the message in ordinary
		time $O(s(n))$.
		
		There is a uniform ordinary program with running time and register
		length $O(s(n))$, together with a total interpretation $m$ of its
		register words, with the following property. For each input length
		$n$, there are commit indices $t_1<\cdots<t_J$ such that, for every
		input $x$ of length $n$, every initial register word $r\in R_n$, and
		every logical index $i$,
		\[
		m(F_{x,i}(r))=
		\begin{cases}
			m(r),         & i<t_1,\\
			g_j(m(r)),    & t_j\leq i<t_{j+1},\quad 1\leq j<J,\\
			g_J(m(r)),    & i\geq t_J.
		\end{cases}
		\]
		Every commit presents its map applied to the original message $m(r)$.
		The final answer also uses that message:
		\[
		\out_x(r)=\gamma(m(r)).
		\]
		For any prescribed constant $d\geq1$, the program can be chosen so
		that $t_1\geq d$ and $t_{j+1}-t_j\geq d$ for every $1\leq j<J$.
	\end{lemma}
	
	\begin{proof}
		Use a register of the form $\pi w_0w_1$, where $\pi$ is a one-cell
		pointer and $w_0,w_1$ are message slots of $k(n)$ cells each, located
		by marking out $k(n)$ cells as in the hypothesis. A pointer value of
		$1$ selects $w_1$, and every other value selects $w_0$. Decode the
		selected slot cell by cell, mapping each symbol to a bit by a fixed
		rule that sends $0$ to $0$ and $1$ to $1$, and ignore all cells
		outside the slots. Every register word therefore decodes to a
		message, including words the program would never write. This
		defines $m(r)$ for every initial register word $r$.
		
		Copy $u=m(r)$ to a work tape. Compute and save $g_1(u),\ldots,g_J(u)$
		and $\gamma(u)$ before writing to either slot. Since $J$ is fixed,
		these computations together take $O(s)$ time and space. For the $j$th
		commit, write a complete encoding of $g_j(u)$ into the unselected slot,
		return the register head to the pointer cell, and switch the pointer
		in one transition. Throughout the write and the return, the pointer
		continues to select the previous message. At the configuration index
		immediately after the switch, the register decodes to $g_j(u)$.
		Repeating this procedure alternates the slots. The program keeps the
		saved bit $\gamma(u)$ in finite control and can enter its corresponding
		halt on the very transition that makes the final pointer switch.
		All earlier control states can be nonaccepting. This proves the
		identities for the decoded messages and the final answer, including
		when the initial register word is malformed.
		
		It remains to schedule the commits at indices that depend only on $n$.
		Before inspecting the register, scan the input to its endmarker and
		prepare a unary block of length $n+1$. Since $\alpha$ is fixed,
		$\alpha$ nested traversals of blocks of this length generate a block
		of length $s(n)=(n+1)^\alpha$ in $O(s(n))$ time, including the return
		scans. Use this procedure to prepare a separate timing tape for each
		stage. All these preparations take the same number of steps on inputs
		of the same length.
		
		Use one stage for decoding and computing the saved values, followed
		by one stage for each slot write. For stage $j$, prepare a timing tape
		with $C_j s(n)$ marked cells, where the fixed integer $C_j$ is large
		enough to cover the stage on every input and register word of that
		length. This allowance includes all head movements. In particular,
		each write stage ends with the register head back at the pointer cell
		and the new pointer value ready in the finite control. Advance the
		timing head one cell per transition. If the work finishes early,
		hold the other heads in place until the timing head reaches its
		endmarker. After each write stage, switch the pointer in one
		transition. The resulting commit indices are the same for every
		$(x,r)$ of input length $n$.
		
		Fixed delays before the first commit and between successive commits
		give the prescribed lower bounds on their spacing. There are only
		constantly many stages, so the total running time and the physical
		register length are both $O(s(n))$.
	\end{proof}
	
	For example, swapping the bits of $(0,1)$ in place can expose $(1,1)$
	after the first write. With two slots, a transmission during the write
	still decodes to $(0,1)$. The pointer switches only when the other
	slot contains $(1,0)$. The phase bit handles what happens after the
	commit: it prevents a later backward departure on the same descent
	from applying a second choice.
	
	\subsection{Proof of Lemma~\ref{lem:gating}}
	
	\begin{proof}
		Use a fixed binary decoding rule for the phase symbol, so every
		register word determines a message $u$ and a phase $e$. Apply
		Lemma~\ref{lem:compiler} to the three maps $g_R,g_0,g_1$, in that
		order, with readout $\gamma(u)$, where
		\[
		g_R(u,e)=(u,0),\qquad
		g_j(u,e)=
		\begin{cases}
			(f_j(u),1), & e=0,\\
			(u,1),      & e=1
		\end{cases}
		\qquad (j=0,1).
		\]
		Choose the commit indices so that $t_R\geq B$, $t_0-t_R\geq 2B$,
		and $t_1-t_0\geq 2B$. The compiler evaluates each map on the pair
		decoded at the start of the selected history. Thus an executed
		backward move preserves that pair below $t_R$, applies $g_R$ on
		$[t_R,t_0)$, applies $g_0$ on $[t_0,t_1)$, and applies $g_1$ at
		indices at least $t_1$, as in Table~\ref{tab:windows}. The register
		length and ordinary running time are both $O(s)$.
		
		Fix any supported pair $+a,-b$. Since $a,b\leq B$, we have
		$t_R\geq b$, and each finite window has width at least $a+b$.
		In particular, a descent by successive $-b$ moves from an index at
		least $t_0$ must enter the reset window: its first index below $t_0$
		lies in $[t_0-b,t_0)\subseteq[t_R,t_0)$. The next $-b$ proposal
		is executed because its departure index is at least $t_R\geq b$.
		
		For~(1), start at any joint state. If its index is below $t_R$, take
		$+a$ proposals until the index first reaches or exceeds $t_R$. This
		places it in $[t_R,t_R+a)\subseteq[t_R,t_0)$ and preserves the
		register. From the resulting index, or directly from any starting
		index at least $t_R$, take $-b$ proposals until the index is below
		$b$. This descent includes a departure from the reset window, which
		sets the phase to zero. Every later departure on the descent preserves
		phase zero, so the final state is ready.
		
		For~(2), start at a ready state with index $i<b$ and message $u$.
		To realize choice $j$, take $+a$ proposals until the index first
		reaches or exceeds $t_j$. The resulting index lies in
		$[t_j,t_j+a)$, inside the chosen window, and the message is still
		ready. One $-b$ proposal now transmits $(f_j(u),1)$. Continue with
		$-b$ proposals until the index is below $b$. Any further departure
		from a choice window preserves the pair because its phase is one.
		A departure from the reset window then restores phase zero, and all
		remaining departures preserve $f_j(u)$. The final state is therefore
		ready with message $f_j(u)$. Each prescribed path is finite and uses
		only supported proposals, so it has positive probability.
		
		For~(3), upward and held proposals preserve the register. On an
		executed backward move, the departure rule either preserves $u$ or
		applies one of $f_0,f_1$. These are exactly the possibilities for
		the projected message transition.
	\end{proof}
	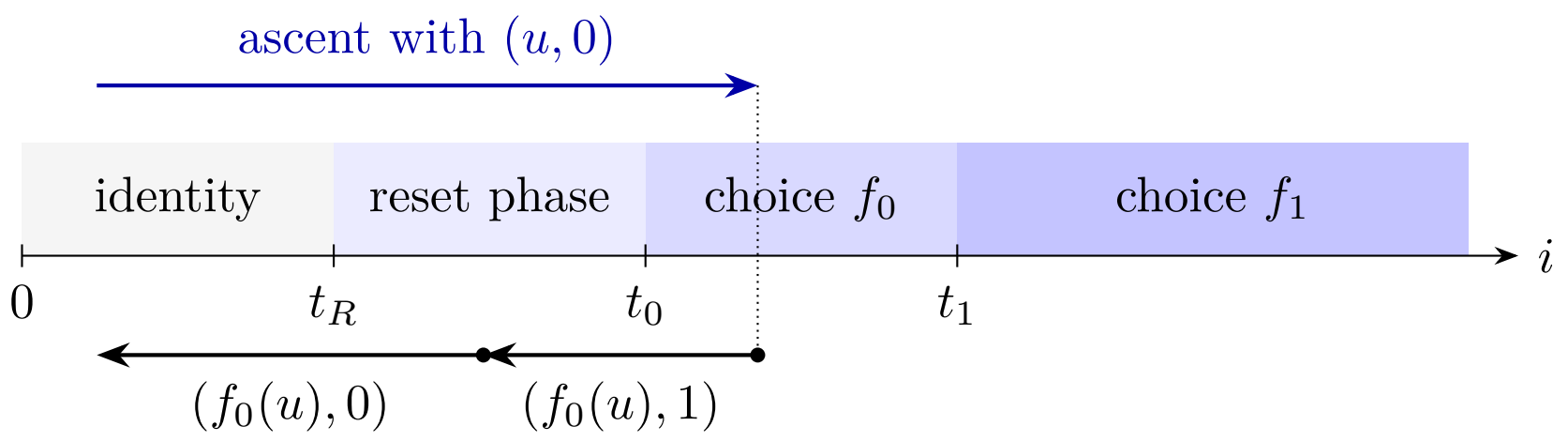
\begin{figure}[htbp]
		\centering
		\begin{tikzpicture}[x=.88cm,y=1cm,>=Stealth,font=\small]
			\fill[gray!8]  (0,0)   rectangle (2.5,.8);
			\fill[blue!8]  (2.5,0) rectangle (5,.8);
			\fill[blue!15] (5,0)   rectangle (7.5,.8);
			\fill[blue!23] (7.5,0) rectangle (11.6,.8);
			
			\draw[->] (0,0) -- (12,0) node[right] {$i$};
			\foreach \pos/\lab in {0/0,2.5/t_R,5/t_0,7.5/t_1}{
				\draw (\pos,.08) -- (\pos,-.08) node[below] {$\lab$};
			}
			\node at (1.25,.4) {identity};
			\node at (3.75,.4) {reset phase};
			\node at (6.25,.4) {choice $f_0$};
			\node at (9.55,.4) {choice $f_1$};
			
			\draw[->,thick,blue!65!black]
			(.6,1.2) -- (5.9,1.2)
			node[midway,above] {ascent with $(u,0)$};
			\draw[densely dotted] (5.9,1.2) -- (5.9,-.7);
			\draw[->,thick] (5.9,-.7) -- (3.7,-.7);
			\draw[->,thick] (3.7,-.7) -- (.6,-.7);
			\fill (5.9,-.7) circle (1.5pt);
			\fill (3.7,-.7) circle (1.5pt);
			\node[below] at (4.8,-.75) {$(f_0(u),1)$};
			\node[below] at (2.15,-.75) {$(f_0(u),0)$};
		\end{tikzpicture}
		\caption{Departure windows and a supported path realizing $f_0$.
			The ascent preserves the ready message $(u,0)$. On the return,
			the first marked departure applies $f_0$ and sets the phase to done.
			The marked departure in the reset window restores the ready phase
			and preserves $f_0(u)$. Arrows represent sequences of proposals.}
		\label{fig:compiler}
	\end{figure}
	
	\section{Stationary existence and the boundary test}
	\label{app:boundary}\label{sec:stability}
	
	This appendix proves Lemmas~\ref{lem:tight} and~\ref{lem:reach}.
	Write $P_x$ for the joint transition kernel on input $x$ and $\E_v$
	for expectation when the chain starts at joint state $v$.
	
	A family of probability laws on a countable state space is
	\emph{uniformly tight} if, for every $\varepsilon>0$, there is a
	finite set $F$ such that $\nu(F)\geq 1-\varepsilon$ for every law
	$\nu$ in the family. The \emph{total variation distance} between two
	laws $\nu$ and $\nu'$ is $\sup_E|\nu(E)-\nu'(E)|$, where the
	supremum ranges over all events. A law has \emph{full support} on a
	countable set if it is concentrated on that set and assigns positive
	probability to each of its states.
	
	\subsection{The exponential estimate and stationary existence}
	
	\begin{proof}[Proof of the bound and existence assertions in Lemma~\ref{lem:tight}]
		Since $K$ is bounded, the function $\theta\mapsto\E e^{\theta K}$ has
		value one and derivative $\E K<0$ at zero. Choose $\theta>0$ such that
		$\lambda=\E e^{\theta K}<1$. Let $P$ be the clock transition kernel,
		and set $V(i)=e^{\theta i}$. The expected value of $V$ after one tick
		from index $i$ is
		\[
		PV(i)=\sum_j P(i,j)V(j).
		\]
		If $i\geq B$, every proposal is executed, so $PV(i)=\lambda V(i)$.
		If $i<B$, the next index is at most $2B$. Thus, with
		$D_{\rm drift}=e^{2\theta B}$, we have
		\[
		PV(i)\leq\lambda V(i)+D_{\rm drift}
		\qquad\text{for every }i\in\N.
		\]
		Iterating this inequality gives
		\[
		\E_i e^{\theta X_t}
		\leq \lambda^t e^{\theta i}
		+D_{\rm drift}\frac{1-\lambda^t}{1-\lambda}
		\leq e^{\theta i}+A,
		\qquad A=\frac{D_{\rm drift}}{1-\lambda}.
		\]
		This proves~\eqref{eq:expbound}. Markov's inequality now gives
		\[
		\Prob_i\{X_t\geq H\}\leq(e^{\theta i}+A)e^{-\theta H}
		\qquad\text{for every }t\geq0.
		\]
		For a fixed input $x$, the register set $R_n$ is finite. Hence, from
		any fixed initial state, the finite sets $\{0,\ldots,H-1\}\times R_n$
		capture arbitrarily large probability uniformly over all ticks.
		
		Let $\nu_t$ be the joint law at tick $t$ from this initial state, and
		let $\bar\nu_N=N^{-1}\sum_{t=0}^{N-1}\nu_t$. These averages are
		uniformly tight. On a countable space, they therefore have a
		subsequence converging in total variation to a probability law
		$\zeta$. Since $\nu_tP_x=\nu_{t+1}$, the averages satisfy
		\[
		\bar\nu_NP_x-\bar\nu_N=\frac{\nu_N-\nu_0}{N}.
		\]
		The total variation distance between $\bar\nu_NP_x$ and $\bar\nu_N$
		is thus at most $1/N$. Applying a transition kernel contracts total
		variation distance, so passing to the subsequence gives
		$\zeta P_x=\zeta$.
		
		If the initial state lies in a closed communicating class, every
		$\nu_t$ is concentrated on that class, as is the limit $\zeta$.
		The chain restricted to the class is irreducible. The return-time
		theorem for countable irreducible chains~\citep[Chapter~5]{Durrett2019}
		then gives positive
		recurrence and, for every state $v$ in the class,
		\[
		\zeta(v)=\frac{1}{\E_v\tau_v^+}>0,
		\]
		where $\tau_v^+$ is the first positive return time to $v$.
	\end{proof}
	
	\subsection{The recurrence criterion and stationary support}
	
	\begin{proof}[Proof of the recurrence criterion in Lemma~\ref{lem:tight}]
		Suppose a boundary state $v$ can reach a state $u$ from which $v$ is
		unreachable. Choose a shortest supported path from $v$ to $u$. With
		positive probability the chain follows this path, which has no
		intermediate visit to $v$, and can never return to $v$ afterward.
		Thus, if $v$ is recurrent, every state reachable from $v$ can return
		to it. In particular, $v$ satisfies~\eqref{eq:boundarycriterion}.
		
		Conversely, suppose $v$ satisfies~\eqref{eq:boundarycriterion}, and
		let $u$ be any state reachable from $v$. Repeated proposals of $-b$
		give a path from $u$ to a boundary state $w$. Since $v\leadsto w$,
		the criterion gives $w\leadsto v$, and hence $u\leadsto v$. The states
		reachable from $v$ therefore form a closed communicating class. By the
		preceding argument, this class is positive recurrent.
		
		To see that a recurrent boundary state exists, partition the finite
		strip $\mathcal B_n$ into groups of mutually reachable states, using
		paths in the full joint chain. Reachability partially orders these
		groups. In a maximal group, every reachable boundary state belongs to
		the same group and can therefore return. Each state in that group
		satisfies~\eqref{eq:boundarycriterion}. Moreover, every recurrent
		class meets the strip: a descent from any state in the class reaches
		the strip and stays in the class because the class is closed.
	\end{proof}
	
	\begin{proof}[Proof of the stationary-support assertion in Lemma~\ref{lem:tight}]
		Let $P_x^t(u,v)$ be the probability that the joint chain is at $v$
		after $t$ ticks when it starts at $u$. If $v$ is transient, the
		expected total number of visits to $v$ is finite from every initial
		state $u$. Indeed, by the strong Markov
		property~\citep[Chapter~5]{Durrett2019}, after the first visit, each return has the same
		probability, strictly less than one, of being followed by another.
		Thus $\sum_{t\geq0}P_x^t(u,v)<\infty$, so $P_x^t(u,v)\to0$ for
		each $u$.
		
		For any stationary law $\zeta$ and every $t$,
		\[
		\zeta(v)=\sum_u\zeta(u)P_x^t(u,v).
		\]
		Each summand is bounded by $\zeta(u)$, and these bounds sum to one.
		Dominated convergence therefore gives $\zeta(v)=0$. Since the state
		space is countable, every stationary law gives total mass zero to the
		transient states.
		
		Conversely, every recurrent state belongs to a closed positive
		recurrent class. The stationary probability of that class gives
		positive mass to each of its states. Extending it by zero outside the
		class gives a stationary law for the entire chain. A wrong bit at any
		recurrent state therefore violates correctness under this law. If
		every recurrent state carries the correct bit, the bit is correct
		almost surely under every stationary law.
	\end{proof}
	
	\subsection{The one-counter reachability bound}
	
	A one-counter system has a finite set of control states and one
	nonnegative integer counter. Each transition changes the counter by
	$-1$, $0$, or $1$. A \emph{guard} specifies whether the transition
	is available when the counter is zero or when it is positive. Every
	transition must leave the counter nonnegative.
	
	Suppose the system has $N$ control states. If a configuration with
	counter value zero can reach another configuration with counter value
	zero, then a path of length at most $14N^2$ connects them
	\citep[Theorem~2.1]{ChistikovEtAl2019}. The counter may become
	positive along this path. We retain the zero and positive guards in
	the encoding below.
	\begin{proof}[Proof of Lemma~\ref{lem:reach}]
		Fix an input $x$ of length $n$, and set $H_{\rm enc}=T+B+1$. We store
		the physical register word $r$ in the finite control. For a clock index
		$i<H_{\rm enc}$, use control $(L_i,r)$ and counter zero. For
		$i\geq H_{\rm enc}$, write
		\[
		i=H_{\rm enc}+Bz+\xi,\qquad z\geq0,\quad 0\leq\xi<B,
		\]
		and use control $(U_\xi,r)$ and counter $z$. This representation is
		unique. The number of controls is
		\[
		N_n=(H_{\rm enc}+B)|R_n|=2^{O(s(n))}.
		\]
		
		In a high control, a supported proposal $k$ changes the residue $\xi$
		to $\xi'$ and produces a carry $\delta$ given by
		\[
		\delta=\left\lfloor\frac{\xi+k}{B}\right\rfloor,\qquad
		\xi'=\xi+k-B\delta.
		\]
		Since $|k|\leq B$, we have $\delta\in\{-1,0,1\}$ and $0\leq\xi'<B$.
		Table~\ref{tab:counter} gives the clock part of each transition.
		
		\begin{table}[htbp]
			\centering\small
			\caption{Clock transitions in the one-counter encoding for a supported
				proposal $k$. The table omits the register component of each control.
				Low controls have transitions only at counter zero.}
			\label{tab:counter}
			\begin{tabular}{@{}lll@{}}
				\toprule
				Source and guard & Condition & Successor and counter\\
				\midrule
				$L_i,\ z=0$   & $i+k<0$                       & $L_i,\ 0$ (hold)\\
				$L_i,\ z=0$   & $0\leq i+k<H_{\rm enc}$       & $L_{i+k},\ 0$\\
				$L_i,\ z=0$   & $i+k\geq H_{\rm enc}$         & $U_{i+k-H_{\rm enc}},\ 0$\\
				$U_\xi,\ z=0$ & $\delta=-1$                   & $L_{H_{\rm enc}+\xi+k},\ 0$\\
				$U_\xi,\ z=0$ & $\delta\in\{0,1\}$            & $U_{\xi'},\ \delta$\\
				$U_\xi,\ z>0$ & $\delta\in\{-1,0,1\}$         & $U_{\xi'},\ z+\delta$\\
				\bottomrule
			\end{tabular}
		\end{table}
		
		To check the transitions across $H_{\rm enc}$, first consider a jump
		from a low index to a high one. Its arrival index satisfies
		$0\leq i+k-H_{\rm enc}<B$, so its high encoding has counter zero.
		In the other direction, a jump from a high index enters the low region
		exactly when $z=0$ and $\delta=-1$. Its arrival index
		$H_{\rm enc}+\xi+k$ then lies between $H_{\rm enc}-B$ and
		$H_{\rm enc}-1$. If instead $z=1$ and $\delta=-1$, the arrival index
		is $H_{\rm enc}+\xi'$, which still has a high control. These rules
		also cover $k=0$ and $|k|=B$.
		
		A held or nonnegative proposal preserves $r$. An executed negative
		proposal from $L_i$ writes $F_{x,i}(r)$. From $U_\xi$, it writes
		$F_{x,T}(r)$ because the departure index exceeds $T$. This rule also
		applies when the move enters the low region. Thus every supported
		proposal produces the encoding of its actual successor, and every
		enabled transition from a valid encoding arises from such a proposal.
		
		Now fix boundary states $v$ and $w$. Both have low controls and
		counter zero. Hence $v\leadsto w$ exactly when the encoding of $v$
		can reach the encoding of $w$. By the zero-to-zero path bound
		\citep[Theorem~2.1]{ChistikovEtAl2019}, whenever such a path exists,
		one exists with at most $14N_n^2$ transitions. Its counter starts at
		zero and increases by at most one per transition, so it never exceeds
		$14N_n^2$.
		
		A nondeterministic machine starts at the encoding of $v$ and accepts
		as soon as it reaches the encoding of $w$, including at step zero.
		At each step it guesses a supported proposal, applies the appropriate
		guard, and computes any required register update by simulating at most
		$T$ ordinary steps. It rejects the branch if the target has not been
		reached after $14N_n^2$ transitions. The control, counter, and step
		count each require $O(\log N_n)=O(s)$ bits, and the simulation uses
		$O(s)$ space. This decides boundary reachability in $\NSPACE(s)$
		with the graph represented implicitly.
	\end{proof}
	\section{Prefix transfer through history replacements}\label{app:observation}
	
	We prove the two locality lemmas used in Theorem~\ref{thm:observation}.
	The first shows that agreement on a prefix survives any number of
	history replacements as long as the clock stays within that prefix.
	
	\begin{lemma}[Register-prefix transfer]\label{lem:register-transfer}
		Let $h\geq0$ be an integer, and let $x,y$ be inputs of length at least
		$h$ that agree on their first $h$ symbols. Start each process at clock
		index zero with its own all-zero register. Give both processes the
		same finite sequence of supported proposals, and suppose the clock
		stays at indices at most $h$.
		
		At every tick, the two processes have the same clock index, their
		persistent registers agree on the first $h$ cells, and their observed
		control states agree. More generally, the two histories selected at
		that tick agree at every logical index from $0$ through $h$ on the
		control state, all head positions, all work-tape contents, and the
		first $h$ register cells. The inputs and registers may have different
		lengths and arbitrary tails beyond their common prefixes.
	\end{lemma}
	
	\begin{proof}
		First fix any two initial register words $r,r'$ that agree on their
		first $h$ cells. Compare the ordinary histories starting from $(x,r)$
		and $(y,r')$. At index zero, the uniform program has the same control
		state, initial head positions, and blank work tapes in both histories.
		Before transition $j+1$, with $0\leq j<h$, the input and register
		heads can be no farther right than cell $j+1\leq h$. The right input
		endmarkers occur at positions $|x|+1$ and $|y|+1$, both at least
		$h+1$. The right register endmarkers also lie beyond cell $h$, since
		each register has length at least its input length plus one. The left
		endmarkers are the same in both histories.
		
		Thus, whenever the histories agree as stated at index $j<h$, the
		heads read the same symbols and the program makes the same writes
		and head movements. Induction gives agreement through index $h$.
		A head can reach cell $h+1$ at index $h$, but its first opportunity
		to read that cell is transition $h+1$. For $h=0$, the comparison
		uses only the initial configurations.
		
		Now induct on clock ticks. At tick zero the blank registers agree on
		their first $h$ cells. The clock indices coincide throughout because
		each clock update depends only on the current index and the common
		proposal. A held or nonnegative proposal preserves the persistent
		registers. For an executed backward move from index $i\leq h$, apply
		the ordinary-history comparison at the departure index: the register
		words $F_{x,i}(r)$ and $F_{y,i}(r')$ agree on their first $h$ cells.
		These become the persistent registers at the next tick.
		
		The prefix agreement therefore holds after every tick. Applying the
		ordinary-history comparison to each newly selected pair of histories
		gives agreement through logical index $h$, in particular at the common
		index where the control state is observed.
	\end{proof}
	
	The next lemma turns a sublinear bound on accepting height into a
	constant bound. The key step is to shorten a member to a prefix while
	preserving the minimum height needed to observe acceptance.
	
	\begin{lemma}[Sublinear accepting height implies a bounded prefix]
		\label{lem:sublinear-accepting-height}
		Consider a uniform program started with a blank register, with the
		sequential access conventions of Section~\ref{sec:model} and register
		length $\ell(n)\geq n+1$. Suppose the program satisfies all-time
		initialized soundness for a language $L$. Let $H\colon\N\to\N$ be
		nondecreasing, with $H(n)/n\to0$ as $n\to\infty$. If every member
		$x\in L$ has a finite supported proposal word that observes acceptance
		from its blank start while the clock stays at indices at most
		$H(|x|)$, then $L\in\RDEF$.
	\end{lemma}
	
	\begin{proof}
		For each $x\in L$, let $h(x)$ be the least height of an accepting
		proposal word from the blank start on $x$, where the height of a word
		is the largest clock index visited along it. The hypothesis guarantees
		that the set of accepting heights is nonempty. Since these heights
		are nonnegative integers, the minimum is attained and satisfies
		\begin{equation}\label{eq:minheight}
			h(x)\leq H(|x|).
		\end{equation}
		
		Fix $x\in L$, write $h=h(x)$, and let $u$ be the prefix of $x$ of
		length $\min\{|x|,h\}$. We claim that $u\in L$ and $h(u)=h$.
		If $u=x$, both assertions are immediate. Otherwise $u$ has length
		$h$ and agrees with $x$ on its first $h$ symbols. By
		Lemma~\ref{lem:register-transfer}, an accepting word of height $h$
		for $x$ also observes acceptance from the blank start on $u$. This
		finite supported word has positive probability, so all-time
		initialized soundness gives $u\in L$. It also gives a witness to
		$h(u)\leq h$. Conversely, a minimizing accepting word for $u$ has
		height at most $h$ and therefore transfers back to $x$. Hence
		$h(x)\leq h(u)$, proving the claim. The same argument covers $h=0$,
		when $u$ is the empty word.
		
		Applying~\eqref{eq:minheight} to $u$ and using $|u|\leq h$ yields
		\[
		h=h(u)\leq H(|u|)\leq H(h).
		\]
		Since $H(m)/m\to0$, choose $h_0\geq1$ such that $H(m)<m$ for every
		$m>h_0$. The displayed inequality forces $h(x)\leq h_0$ for every
		member $x$.
		
		Now let $x$ and $y$ have length at least $h_0$ and agree on their
		first $h_0$ symbols. If $x\in L$, an accepting word of height at most
		$h_0$ transfers from $x$ to $y$. All-time initialized soundness then
		gives $y\in L$. The same implication holds with $x$ and $y$
		interchanged. Thus the first $h_0$ symbols determine membership for
		all words of length at least $h_0$, as required.
	\end{proof}

\section{Operational-cost proofs}\label{sec:costproofs}

Throughout, we use the clock, departure height $h$, update target $M$
and erasure rule of Section~\ref{sec:costs}.

\begin{proof}[Proof of Theorem~\ref{thm:waiting}]
For a fixed proposal, the new clock index is a nondecreasing function
of the old one. Use the same proposals for a clock started at zero and
one started with the stationary law $\varpi$. The first clock stays
at or below the second, so $\Prob_0(X_j\geq h)\leq\rho^h$ for every $j$.
The next proposal is independent of $X_j$, giving
$\E_0 U_t\leq q t\rho^h$. Markov's inequality gives
\eqref{eq:waitinglower} and the quantile lower bound.

For the mean upper bound, let $d_i$ be the mean time to hit $i+1$
from $i$. A block of $i+1$ upward proposals reaches the target from
any state at most $i$, so disjoint blocks give a geometric upper bound
and $d_i$ is finite. At zero, the hold gives $d_0=1/p$. At $i\geq1$,
a downward proposal incurs the wait to return to $i$ and then the
wait to hit $i+1$. The first-step equation $d_i=1+q(d_{i-1}+d_i)$
gives $d_i=(\rho^{-i-1}-1)/(q-p)$.
The time $\sigma_h$ to first hit $h$ from zero therefore has mean
\begin{equation}\label{eq:hittingexact}
 \E_0 \sigma_h=\sum_{i=0}^{h-1}d_i
 =\frac{q}{(q-p)^2}(\rho^{-h}-1)-\frac{h}{q-p}.
\end{equation}
After hitting $h$, the first negative proposal gives a useful update and
has mean waiting time $1/q$. Immediately after each useful update the
index is at least $h-1$. The monotone coupling and the strong Markov
property~\citep[Section~5.2]{Durrett2019} bound the mean time to the
next update by $d_{h-1}+1/q$. Hence
\[
 \E_0\tau_M\leq\E_0 \sigma_h+\frac1q
 +(M-1)\left(d_{h-1}+\frac1q\right).
\]
Both $\E_0\sigma_h+1/q$ and $d_{h-1}+1/q$ are at most
$C(p)\rho^{-h}$, since $1/(q-p)\leq q/(q-p)^2$ and $\rho^{-h}\geq1$.
This proves~\eqref{eq:waitingmean}. Finally,
$\Prob_0(\tau_M>t)\leq C(p)M\rho^{-h}/t$ gives the quantile upper bound.
\end{proof}

\begin{proof}[Proof of Theorem~\ref{thm:erasure}]
The clock still satisfies the exponential bound of Lemma~\ref{lem:tight},
and the counter set is finite. The Ces\`aro argument used there gives
existence. Every stationary joint law has clock marginal $\varpi$,
the unique stationary law of the autonomous clock.

For necessity, on a tick without an erasure the counter increments
by $I_t\one\{Y_t<M\}$, and on an erasure tick the counter drops by
$Y_t$. Stationarity and independence of the erasure give
\[
 0=-\varepsilon\E Y_t+(1-\varepsilon)\E[I_t\one\{Y_t<M\}]
 \leq-\varepsilon\E Y_t+(1-\varepsilon)q\rho^h.
\]
Since $M\Prob(Y_t=M)\leq\E Y_t$, this proves~\eqref{eq:erasurenec}.
Rearranging it at probability $\beta$ gives~\eqref{eq:necessaryeps}.

For sufficiency, start in a stationary joint law and look at the next
$t$ ticks. If they contain no erasure and at least $M$ useful updates,
the final counter is $M$ for every initial counter value. The union
bound gives probability at most $\varepsilon t$ of an erasure in the
block. Use the same proposals for the stationary clock and a clock
started at zero. The stationary clock gives at least as many useful
updates, since at a fixed proposal the update indicator is
nondecreasing in the departure index. Theorem~\ref{thm:waiting} gives
\[
 \Prob(Y=M)\geq1-\varepsilon t-\frac{C(p)M\rho^{-h}}{t}.
\]
Let $W=M\rho^{-h}\geq1$ and choose
$t=\ceil{2C(p)W/(1-\beta)}\leq D(p,\beta)W$.
The last term of this bound is at most $(1-\beta)/2$. The stated
bound on $\varepsilon$ makes $\varepsilon t$ at most the same number.
This proves sufficiency, including $\varepsilon=0$.
\end{proof}

\bibliographystyle{elsarticle-num}
\bibliography{references}

@article{Deutsch1991,
  author = {Deutsch, D.},
  title = {Quantum mechanics near closed timelike lines},
  journal = {Physical Review D},
  year = {1991},
  volume = {44},
  number = {10},
  pages = {3197--3217},
  doi = {10.1103/PhysRevD.44.3197}
}

@article{AaronsonWatrous2009,
  author = {Aaronson, S. and Watrous, J.},
  title = {Closed timelike curves make quantum and classical computing equivalent},
  journal = {Proceedings of the Royal Society A},
  year = {2009},
  volume = {465},
  number = {2102},
  pages = {631--647},
  doi = {10.1098/rspa.2008.0350}
}

@misc{SayYakaryilmaz2011Narrow,
  author = {Say, A. C. C. and Yakary{\i}lmaz, A.},
  title = {Computation with narrow {CTCs}},
  year = {2011},
  howpublished = {arXiv:1102.1199v2},
  note = {Preprint version},
  url = {https://arxiv.org/pdf/1102.1199v2}
}

@inproceedings{ODonnellSay2014,
  author = {O'Donnell, R. and Say, A. C. C.},
  title = {One time-traveling bit is as good as logarithmically many},
  booktitle = {34th International Conference on Foundation of Software Technology and Theoretical Computer Science (FSTTCS 2014)},
  series = {Leibniz International Proceedings in Informatics},
  volume = {29},
  publisher = {Schloss Dagstuhl--Leibniz-Zentrum f{\"u}r Informatik},
  year = {2014},
  pages = {469--480},
  doi = {10.4230/LIPIcs.FSTTCS.2014.469}
}

@misc{AaronsonEtAl2024,
  author = {Aaronson, S. and Bavarian, M. and Cubitt, T. and Grewal, S. and Gueltrini, G. and O'Donnell, R. and Raat, M.},
  title = {Computability theory of closed timelike curves},
  year = {2024},
  howpublished = {arXiv:1609.05507v2},
  note = {Corrected version},
  url = {https://arxiv.org/abs/1609.05507v2}
}

@misc{Didehvar2023,
  author = {Didehvar, F.},
  title = {Theory of fuzzy time computation ({TC}$^{*}$ vs {TC} \& {TQC})},
  year = {2023},
  howpublished = {SSRN 4572734},
  note = {Manuscript dated 1 January 2023},
  doi = {10.2139/ssrn.4572734}
}

@misc{Didehvar2026,
  author = {Didehvar, F.},
  title = {Theory of fuzzy time computation ({TC}$^{*}$ vs {TC}, edited version)},
  year = {2026},
  howpublished = {PhilArchive preprint DIDTOF-2},
  url = {https://philpapers.org/archive/DIDTOF-2.pdf}
}

@misc{DidehvarFarhat2026RandomClocks,
  author = {Farhat, S. and Didehvar, F.},
  title = {Random clocks on computation paths: How far can they reach, and when does stepping back help?},
  year = {2026},
  howpublished = {arXiv:2610.25830v2},
  note = {Preprint},
  url = {https://arxiv.org/abs/2610.25830v2}
}

@article{Kuroda1964,
  author = {Kuroda, S.-Y.},
  title = {Classes of languages and linear-bounded automata},
  journal = {Information and Control},
  year = {1964},
  volume = {7},
  number = {2},
  pages = {207--223},
  doi = {10.1016/S0019-9958(64)90120-2}
}

@article{ChistikovEtAl2019,
  author = {Chistikov, D. and Czerwi{\'n}ski, W. and Hofman, P. and Pilipczuk, M. and Wehar, M.},
  title = {Shortest paths in one-counter systems},
  journal = {Logical Methods in Computer Science},
  year = {2019},
  volume = {15},
  number = {1},
  pages = {19:1--19:28},
  doi = {10.23638/LMCS-15(1:19)2019}
}

@inproceedings{PaulPippengerSzemerediTrotter1983,
  author = {Paul, W. J. and Pippenger, N. and Szemer{\'e}di, E. and Trotter, W. T.},
  title = {On determinism versus non-determinism and related problems (preliminary version)},
  booktitle = {24th Annual Symposium on Foundations of Computer Science},
  publisher = {IEEE},
  year = {1983},
  pages = {429--438},
  doi = {10.1109/SFCS.1983.39}
}

@article{Immerman1988,
  author = {Immerman, N.},
  title = {Nondeterministic space is closed under complementation},
  journal = {SIAM Journal on Computing},
  year = {1988},
  volume = {17},
  number = {5},
  pages = {935--938},
  doi = {10.1137/0217058}
}

@article{Szelepcsenyi1988,
  author = {Szelepcs{\'e}nyi, R.},
  title = {The method of forced enumeration for nondeterministic automata},
  journal = {Acta Informatica},
  year = {1988},
  volume = {26},
  number = {3},
  pages = {279--284},
  doi = {10.1007/BF00299636}
}

@article{HopcroftPaulValiant1977,
  author = {Hopcroft, J. E. and Paul, W. J. and Valiant, L. G.},
  title = {On time versus space},
  journal = {Journal of the ACM},
  year = {1977},
  volume = {24},
  number = {2},
  pages = {332--337},
  doi = {10.1145/322003.322015}
}

@book{AroraBarak2009,
  author = {Arora, S. and Barak, B.},
  title = {Computational Complexity: A Modern Approach},
  publisher = {Cambridge University Press},
  year = {2009},
  doi = {10.1017/CBO9780511804090}
}

@article{Ibarra1972,
  author = {Ibarra, O. H.},
  title = {A note concerning nondeterministic tape complexities},
  journal = {Journal of the ACM},
  year = {1972},
  volume = {19},
  number = {4},
  pages = {608--612},
  doi = {10.1145/321724.321727}
}

@article{Seiferas1977,
  author = {Seiferas, J. I.},
  title = {Relating refined space complexity classes},
  journal = {Journal of Computer and System Sciences},
  year = {1977},
  volume = {14},
  number = {1},
  pages = {100--129},
  doi = {10.1016/S0022-0000(77)80006-X}
}

@article{Savitch1970,
  author = {Savitch, W. J.},
  title = {Relationships between nondeterministic and deterministic tape complexities},
  journal = {Journal of Computer and System Sciences},
  year = {1970},
  volume = {4},
  number = {2},
  pages = {177--192},
  doi = {10.1016/S0022-0000(70)90006-X}
}

@article{BrzozowskiLiLiu2012,
  author = {Brzozowski, J. and Li, B. and Liu, D.},
  title = {Syntactic complexities of six classes of star-free languages},
  journal = {Journal of Automata, Languages and Combinatorics},
  year = {2012},
  volume = {17},
  number = {2--4},
  pages = {83--105},
  doi = {10.25596/jalc-2012-083}
}

@article{PerlesRabinShamir1963,
  author = {Perles, M. A. and Rabin, M. O. and Shamir, E.},
  title = {The theory of definite automata},
  journal = {IEEE Transactions on Electronic Computers},
  year = {1963},
  volume = {EC-12},
  number = {3},
  pages = {233--243},
  doi = {10.1109/PGEC.1963.263534}
}

@inproceedings{Santhanam2001,
  author = {Santhanam, R.},
  title = {On separators, segregators and time versus space},
  booktitle = {16th Annual IEEE Conference on Computational Complexity},
  publisher = {IEEE},
  year = {2001},
  pages = {286--294},
  doi = {10.1109/CCC.2001.933895}
}

@article{ElnozahyEtAl2002,
  author = {Elnozahy, E. N. and Alvisi, L. and Wang, Y.-M. and Johnson, D. B.},
  title = {A survey of rollback-recovery protocols in message-passing systems},
  journal = {ACM Computing Surveys},
  year = {2002},
  volume = {34},
  number = {3},
  pages = {375--408},
  doi = {10.1145/568522.568525}
}

@article{EvansMajumdar2011,
  author = {Evans, M. R. and Majumdar, S. N.},
  title = {Diffusion with stochastic resetting},
  journal = {Physical Review Letters},
  year = {2011},
  volume = {106},
  number = {16},
  pages = {160601},
  doi = {10.1103/PhysRevLett.106.160601}
}

@book{Durrett2019,
  author = {Durrett, R.},
  title = {Probability: Theory and Examples},
  edition = {Fifth},
  publisher = {Cambridge University Press},
  year = {2019},
  doi = {10.1017/9781108591034}
}
\end{document}